\documentstyle[aps]{revtex}
\input psfig.sty

\def\k{{\bf k}}

\def\q{{\bf q}}

\def\R{{\bf R}}
\def\r{{\bf r}}

\def\vs{{\bf v}_s}
\def\v{{\bf v}}

\def\A{{\bf A}}

\def\Z{{\cal Z}}
\def\D{{\cal D}}
\def\D{{\cal D}}
\def\Dd{{\bar{\cal D}}}

\def\1v{{\bf v}}
\def\dx2-y2{d_{x^2-y^2}}
\def\dxy{d_{xy}}

\def\ltsim{\vbox {\hbox{\lower .8\baselineskip \hbox{$<$}} \break
                 \hbox{\lower 0.2\baselineskip \hbox{$\sim$}} } }

\begin{document}

\title{Vortex in a $d$-wave Superconductor at Low Temperatures}

\author{Mei-Rong Li,$^{a,b}$ P. J. Hirschfeld$^{c}$ and P. W\"olfle$^a$}

\address{
$^a$Institut f\"ur Theorie der Kondensierten Materie, Universit\"at
Karlsruhe, 76128 Karlsruhe, Germany\\
$^b$Department of Physics, Nanjing University, Nanjing 210093, P.R.China\\
$^c$Department of Physics, University of Florida, Gainesville, FL 32611}

\maketitle

\begin{abstract}

A systematic perturbation theory is developed to describe the magnetic 
field-induced subdominant $s$- and $d_{xy}$-wave order parameters in the 
mixed state of a $d_{x^2-y^2}$-wave superconductor, enabling us to obtain, 
within weak-coupling BCS theory, analytic results for the free energy of 
a $d$-wave superconductor in an applied magnetic field $H_{c1}\ltsim H\ll 
H_{c2}$ from $T_c$ down to very low temperatures. Known results for a single 
isolated vortex in the Ginzburg-Landau regime are recovered, and the behavior 
at low temperatures for the subdominant component is shown to be qualitatively 
different. In the case of subdominant $d_{xy}$ pair component, superfluid 
velocity gradients and an orbital Zeeman effect are shown to compete in
determining the vortex state, but for realistic field strengths the latter 
appears to be irrelevant. On this basis, we argue that recent predictions of 
a low-temperature phase transition in connection with recent thermal 
conductivity measurements are unlikely to be correct.

\end{abstract}

\section{INTRODUCTION}

Vortices in classic superconductors involve a winding of $2\pi$ of the order 
parameter phase $\phi(\r)$ around the vortex line. Since the order parameter 
$\Delta_\k=|\Delta_\k|e^{i\phi}$ in a $d_{x^2-y^2}$-wave superconductor like 
the high-$T_c$ cuprates is also simply a complex scalar with single global 
phase $\phi$, it was initially expected that the vortex state in the cuprates
might be {\it structurally} quite similar to the textbook case. One remarkable 
difference was pointed out by Joynt\cite{Joynt}: in the $d$-wave case, if 
subdominant pair potential components of different symmetry exist, corresponding 
order parameter components can be induced at $T_c$ by any probe which couples 
to gradients of the order parameter, even if the subdominant zero field ``bare'' 
critical temperatures are very small or zero. Thus the structure of a $d$-wave 
vortex generically involves admixtures of different symmetry order parameters. 
Secondly, as noted by Volovik\cite{volovik1}, the traditional roles played by  
extended and localized quasiparticle states in classic superconductors are 
reversed in the $d$-wave case. In classic superconductors at low temperatures, 
Caroli-de Gennes-Matricon bound states in the vortex core \cite{Caroli} dominate 
the electronic density of states because extended states are fully gapped and 
therefore unoccupied. In the $d$-wave case, the existence of order parameter 
nodes inhibits the formation of the bound states (their very existence is
questionable \cite{FranzTesanovic,Kita}), and populates the extended ones, which 
are found to dominate thermodynamics at low temperatures and magnetic fields.

\vskip .2cm
Several authors \cite{Soininen,RXT,Berlinsky,Franz,Heeb,Ichioka,Koyama,WWL}
have attacked the $d$-wave mixed state structure problem in recent years, armed 
with these ideas. Early studies focussed on an isolated $d$-wave vortex, allowing 
for an induced $s$-wave order parameter component and solving the Bogoliubov-de 
Gennes equations on a lattice.\cite{Soininen} Within the Ginzburg-Landau (GL) 
theory, similar results were obtained.\cite{RXT,Berlinsky,Franz,Heeb}
The $s$ component was shown to have opposite winding number to that of the parent
$d_{x^2-y^2}$ in the core regions. Far away from the core center, it decays as 
$1/r^2$, and its winding number becomes 3,\cite{Soininen,RXT,Berlinsky,Franz,Heeb}, 
implying that there are four extra vortices in the  $s$-field at large distances 
from the main core. \cite{Berlinsky,Franz,Heeb} These results were confirmed
by numerical solutions of the Eilenberger equations by Ichioka et al.\cite{Ichioka}
The possibility of an induced $d_{xy}$ component was also allowed for in Ref.  
\cite{Ichioka}, which concluded that the structure in this case was similar to 
that of the (induced) $s$-wave case, except that the induced order parameter
at large distances was found to decay more rapidly, roughly as $1/r^4$, and have 
opposite winding number 5. Koyama and Tachiki\cite{Koyama} pointed out, however, 
that if the calculation is done in a gauge invariant manner there is an additional 
term in the free energy not found by Ichioka et al., involving a Zeeman coupling 
of the field to an intrinsic orbital magnetic moment in a state with structure 
close to a uniform $d_{x^2-y^2}+i d_{xy}$. They furthermore showed, within a GL 
framework with coefficients determined by BCS weak-coupling theory, that this 
term is proportional to the  particle-hole asymmetry of the normal metal from 
which the superconductor condenses, and dominates sufficiently far from the 
vortex core.

\vskip .2cm
Interest in the possibility of order-parameter mixing in the vortex state was 
heightened by the experimental observation \cite{Krishana,Aubin,Ando} of a 
plateau in the thermal conductivity as a function of the magnetic field
$H$ when $H$ is above some critical value $H^*$. Krishana et al.,\cite{Krishana} 
in particular, speculated that their observation of a sharp kink at $H^*$ might 
be explained by the sudden onset of an out-of-phase $d_{xy}$ component at this 
critical field. The new high-field state was proposed to be fully gapped, with 
vanishing  quasiparticle transport. There are several difficulties with this 
explanation, which we discuss below, but theorists were nonetheless persuaded to
revisit the problem.  

\vskip .2cm
Laughlin \cite{Laughlin} then pointed out, in analogy to the quantum Hall state,  
the peculiar nature of the time-reversal symmetry breaking $d_{x^2-y^2}+id_{xy}$ 
state, which appears to be quite different from possible ground states in the 
case of the $d,s$-wave mixture.  He proposed that the development of a magnetic 
moment coupling to the magnetic field might account for the phase transition,
and put forward a $T$- and $H$-dependent free energy functional driving the phase 
transition, and found a critical field $H^*\sim T^2$ similar to experiment. This 
special free energy functional does not have a microscopic basis and it is not 
clear yet whether there is such a field-induced secondary phase transition.  
To date, no magnetic-field induced transition has been found in relevant numerical 
studies of the vortex lattice.\cite{Ichioka,Kita}

\vskip .2cm
Laughlin's argument ignored the physics of the core region, thought to be 
negligible, but Ramakrishnan pointed out that quite similar effects are to be 
expected due to a combination of superfluid velocity Doppler shifts and Andreev 
reflection near the vortex cores.\cite{Ramakrishnan} In higher fields, he proposed 
that these local $d_{x^2-y^2}+id_{xy}$ patches might overlap, causing a transition 
to a uniform gapped state. This scenario is similar to one proposed by Movshovich 
et al.\cite{Movshovich} in the related case of magnetic impurities in a $d$-wave 
superconductor in zero field. Finally, Balatsky \cite{Balatsky} has recently 
investigated the effect of the orbital Zeeman term near the upper critical field, 
and argued that the $d$-wave state is always unstable to a $d_{x^2-y^2}+id_{xy}$ 
mixture. An unusual collective clapping mode in association with the relative 
phase of $d_{xy}$ to $d_{x^2-y^2}$ was predicted in this superconductor. 
\cite{BalatskyKumar}

\vskip .2cm
Several important questions have not been addressed in the analyses of these 
issues thus far  and have motivated this work. First and foremost, we would 
like to  understand whether a phase transition of the type proposed by Krishana 
et al. is possible. Analyses in the GL regime are not applicable, and numerical
calculations\cite{Kita,Soininen,Franz,Ichioka} are not always useful to understand 
competing physical effects. We have therefore developed a systematic calculational 
approach capable of treating carefully the relevant quasiparticle states in the 
presence of spatially varying superflow together with the relevant subdominant 
order parameter components on an equal footing in the low-temperature phase.  
Our theory works for $H\ll H_{c2}$, in which case the vortex core region can be 
safely neglected. Secondly, we would like to understand the structure of the 
vortex state and the role of the quasiparticles as a preliminary to the 
yet-unsolved problem of quasiparticle transport in applied magnetic field. In
addition to being inapplicable at low temperatures, the GL calculations on which  
most of one's intuition for this problem is based are unable to predict magnitudes 
of physical effects since they are based entirely on symmetry considerations. 
This is particularly important in the case of the orbital Zeeman coupling in the 
$d_{x^2-y^2},d_{xy}$ mixing problem.  The magnitude of the induced orbital moment
is a very difficult quantity to estimate properly, as one might {\it a priori} 
deduce by analogy to the intrinsic orbital angular momentum  problem in the 
$^3He$ A-phase. In this case, it was found that naive calculations dramatically
overestimated this effect, and we show here  in fact that the orbital Zeeman 
coupling in the current problem is quite small. 

\vskip .2cm
In this paper, we adopt a semiclassical approach, expanding the BCS free energy 
in powers of the local superfluid velocity, local order parameter magnitude 
fluctuations, and their gradients. We thus neglect states possibly localized in 
the vortex core, and other quasiparticle bandstructure effects in a periodic 
vortex lattice discussed recently by several authors 
\cite{Anderson,GorkovSchrieffer,FranzTes,Halperin}. Initially Volovik  
\cite{volovik1} and Kopnin and Volovik \cite{KopninVolovik} proposed that a 
semiclassical analysis of this type should be valid only down to a scale 
$(\Delta^2_0/E_F)\sqrt{H/H_{c2}}$. Recent numerical work \cite{FranzTes,Halperin} 
indicated, however, that the true crossover scale is much smaller for realistic 
systems with $\Delta_0/E_F\ll 1$. Such fine details of the true quantum 
quasiparticle band structure will also be smeared out by impurity effects. We 
believe, therefore, that our neglect of the vortex core and quasiparticle
bandstructure will be justified for cuprate superconductors at low fields 
($H\ll H_{c2}$) and temperatures, and that the current analysis will thus be
adequate. 

\vskip .2cm
We begin by presenting in Sec. II the method, which involves a functional 
integral representation of the BCS free energy $F$, which we then expand in
powers of slow superfluid velocity gradients and small subdominant order 
parameter components. Analytical results for $F$ in the GL regime, the low 
temperature regime, and, for $s$-wave case, an ultralow temperature regime 
where nonlinear superflow effects dominate, are given. This allows us in 
Sec. III to calculate the order parameter fluctuations directly. We then apply 
these results to the comparison of structure of a single isolated vortex with 
$s$ or $d_{xy}$ subdominant pairing at various temperatures in Sec. IV, and go 
on in Sec. V to discuss the prospects for observing a low temperature field-induced
transition of this structure. In Sec. VI we discuss existing experiments
and make some comments on the various available scenarios. In Appendix A, a 
detailed derivation of the free energy is presented, while Appendix B is devoted 
to a general calculation of the spontaneous magnetization in a 
$d_{x^2-y^2}+id_{xy}$-wave superconductor.

\section{FREE ENERGY} 

We start from a two-dimensional (2D) phenomenological BCS mean-field Hamiltonian 
in the mixed state: \cite{LHWprb}
\begin{eqnarray}
{\widehat H}_{\rm MF}&=&\sum_\sigma\int d^2\r c^\dagger_{\sigma}(\r)
\left \{ {1\over 2m} \left [-i{\vec \nabla} - {e\over c}\A(\r)\right ] ^2-\mu 
\right \} c_{\sigma}(\r) +  \int d^2\r d^2\r' \left [\Delta(\r,\r')
c^\dagger_\uparrow(\r) c^\dagger_\downarrow(\r')+h.c.\right ] \nonumber\\
&& - \int d^2\r d^2\r' V(\r-\r')|b(\r,\r')|^2,   \label{ham}
\end{eqnarray}
where 
\begin{eqnarray}
V(\r)= V_d \Phi_d(\r) +V_s + V_{xy} \Phi_{xy}(\r) 
\end{eqnarray}
 with 
$\Phi_i$ characterizing the irreducible representations $d_{x^2-y^2}$, $s$, 
and $d_{xy}$, for which the lowest order basis functions over a circular Fermi 
surface are
\begin{eqnarray}
\Phi_{i\k}=  \left \{
\begin{array}{ll}
\cos 2 \varphi, & \;\;\;\;\; i=d_{x^2-y^2} \\
1, & \;\;\;\;\; i=s \\
\sin 2 \varphi, & \;\;\;\;\; i=d_{xy}
\end{array}
\right.  ,
\end{eqnarray}
$\Delta (\r,\r')= V(\r-\r')b(\r,\r')$ is the pairing order parameter with 
$b(\r,\r')=\langle c_\downarrow (\r') c_\uparrow (\r)\rangle$, and $\A(\r)$ 
the magnetic vector potential. Throughout the paper $\hbar=k_B=1$ units are 
chosen. The magnetic field in the problem is perpendicular to the 2D plane, 
i.e., along the $\hat{z}$ direction. We assume that $V_d<0$, and $V_d, V_s$ 
and $V_{xy}$ take such values that in the absence of the magnetic field, the 
superconducting state is of $d_{x^2-y^2}$-wave symmetry with $\Delta (\r,\r')
=\Delta_0  \Phi_d(\r-\r')$. In the mixed state, $\Delta(\r,\r')$ takes the 
form of $\Delta(\r,\r')=e^{i\phi({\r+\r' \over 2})}\widetilde\Delta({\r+\r' 
\over 2},\r-\r')$, where $\widetilde\Delta(\R,{\vec \rho})=\bar{\Delta}_d(\R)
\Phi_d({\vec \rho})+\D_s(\R) + \D_{xy}(\R)\Phi_{xy}({\vec \rho}) $, with 
$\bar{\Delta}_d(\R)=\Delta_0+\D_d(\R)$. The  $\D_i(\r)$ are the magnetic 
field-induced pairing order parameter deviations from their values in zero field.
  
The partition function of Hamiltonian (\ref{ham}), after making a canonical 
transformation \cite{gaugetran} to eliminate the phase field of the pairing 
order parameter $\phi(\r)$, \cite{LHWprb} is:
\begin{eqnarray}
&&{\cal Z}= \Z_0 {\rm exp} \left \{
\sum_n {\rm Tr} \, {\rm ln}  
\left [ \widehat{M}_0 + \left (
 \begin{array}{cc}
\widehat{V}_1  &  \widehat{\D}  \\
 \widehat{\D}^\dagger   &  \widehat{V}_2 
  \end{array}
\right ) \right ] \right \}, 
\label{Z1}
\end{eqnarray}
where 
\begin{eqnarray}
\Z_0&=&\exp (-F_0/T), \nonumber  \\
F_0&=& -\int d^2\r \left ({|\bar{\Delta}_d(\r)|^2 \over V_d} 
+ {|\D_s(\r)|^2\over V_s} + {|\D_{xy}(\r)|^2\over V_{xy}}
\right ) ,      \label{f0}
\end{eqnarray}
and, in the momentum representation,
\begin{eqnarray}
(\widehat{M}_0 )_{\k,\k'} = \left (  
\begin{array}{cc}
-i \omega_n + \epsilon_\k   &  \Delta_\k  \\
\Delta_\k  & -i \omega_n - \epsilon_\k 
\end{array}
\right ) \delta_{\k,\k'}
\end{eqnarray}
\begin{eqnarray}
(\widehat{V}_1)_{\k,\k'}&=& \int d^2\r e^{i(\k'-\k)\cdot \r} 
 \left [-\v_s(\r) \cdot \k' + {1\over 2} m v^2_s(\r)  \right ]  , \label{v1} \\
(\widehat{V}_2)_{\k,\k'}&=& \int d^2\r e^{i(\k'-\k)\cdot \r} 
 \left [-\v_s(\r) \cdot \k - {1\over 2} m v^2_s(\r)  \right ]  , \label{v2} \\
\widehat{\D}_{\k,\k'}&=& \sum_i (\widehat{\D}_i)_{\k,\k'}= \sum_i
 \int d^2\r e^{i(\k'-\k)\cdot \r} \D_i (\r) \Phi_{i{\k+\k' \over 2}}
\end{eqnarray}
with $\omega_n=(2n+1)\pi T $ the fermion Matsubara frequency, $\Delta_\k=
\Delta_0 \Phi_{d\k}$, and 
\begin{eqnarray}
\v_s(\r)={e\A(\r)\over mc}-{\nabla\phi(\r)\over 2m} \label{vs1}
\end{eqnarray}
the supercurrent velocity. In writing down Eq. (\ref{Z1}) we have used
the relation ${\overrightarrow \nabla} \cdot \v_s(\r)=0 $ which corresponds 
to the conservation of the supercurrent. Otherwise, $m v^2_s(\r)/2$
in Eqs. (\ref{v1}) and (\ref{v2}) has no significant effect and will be neglected 
hereafter.   

\vskip .2cm
The free energy resulting from Eq. (\ref{Z1}) is
 \begin{eqnarray}
F&=&-T\,{\rm Tr} \, {\rm ln} \, {\cal Z} 
=F_0 -T\,{\rm Tr} \, {\rm ln} \, \widehat{M}_0
+T\, \sum^\infty_{m=1} {1\over m} \,{\rm Tr}\, \hat{p}^m ,
\label{fe1}
\end{eqnarray}
where 
\begin{eqnarray}
\hat{p}=\hat{g}  \cdot \left (
\begin{array}{cc}
\widehat{V}_1  &  \widehat{\D}  \\
  \widehat{\D}^\dagger   &   \widehat{V}_2
  \end{array}
\right ) ,  
\end{eqnarray}  
with $\hat{g}$ the Green-function matrix which, in the momentum representation, 
is
\begin{eqnarray}
\hat{g}_\k=\left (
\begin{array}{cc}
g_{1\k}  & g_{2\k}  \\
g_{2\k}  & g_{4\k} 
  \end{array}
\right ) 
=-\left (
\begin{array}{cc}
{i\omega_n+\epsilon_{\k} \over W_{n\k}}  &  {\Delta_\k \over W_{n\k}}  \\  
{\Delta_\k \over W_{n\k}}  & {i\omega_n-\epsilon_{\k} \over W_{n\k}}
  \end{array}
\right ) .   \label{gf}
\end{eqnarray}
Here,
\begin{eqnarray}
W_{n\k}=\omega^2_n+E^2_\k, \;\;\;\;\;\;\;\; E_\k=\sqrt{\epsilon_\k^2+\Delta^2_k} .
\end{eqnarray}
The calculation of the trace of $\hat{p}^m$ in Eq. (\ref{fe1}) can be done by 
noting that 
\begin{eqnarray}
{\rm Tr}\, \hat{p}^m &=& \prod^m_{j=1} \int {d^2\k_j \over (2\pi)^2}  
\int d^2\r_j  e^{i(\k_j-\k_{j+1})\cdot \r_j}\;
{\rm Tr} \left (\prod^m_{j=1}  \hat{p}_{\k_j,\k_{j+1}}(\r_j) \right ) ,
\label{trpm}
\end{eqnarray}
where $\k_{m+1}=\k_1$. In most of the bulk region not close to a vortex core,
$\v_s$ as well as the  $\v_s$-induced $\D_i$ are spatially slowly varying 
functions. Thus we are allowed to expand $\v_s(\r_j)$ and $\D_i(\r_j)$ in 
$\hat{p}_{\k,\k_1}(\r)$ in Eq. (\ref{trpm}) as power series in their derivatives 
\cite{RXT},
\begin{eqnarray}
\v_s(\r_j) &\simeq &  e^{(\r_j-\r)\cdot {\vec \nabla_\r }} \v_s(\r) 
= \v_s(\r)+[(\r_j-\r)\cdot {\vec \nabla_\r } ] \v_s(\r) + \cdots , \nonumber \\
\D_i(\r_j) &\simeq &  e^{(\r_j-\r)\cdot {\vec \nabla_\r } } \D_i(\r) 
= \D_i(\r)+[(\r_j-\r)\cdot {\vec \nabla_\r } ] \D_i(\r) + \cdots , 
\label{vsexpansion}
\end{eqnarray}
the first few terms of which make main contribution to Tr$\hat{p}^m$. 
This property enables us to develop a perturbation theory to obtain
the free energy. The resulting calculation is straightforward but tedious, 
and is summarized in Appendix A. The final result for the free energy with 
respect to the pairing order parameters $\widetilde{F}=\int d^2\r \tilde{f}(\r)$, 
where the free energy density $\tilde{f}(\r)$, keeping terms up to quadratic
in $\D_i$, is   
\begin{eqnarray}
&& \;\;\;\;\;\;\;\; \tilde{f}(\r) = f_{\bar{\Delta}_d}(\r)+ 
f_s(\r)+f_{xy}(\r)+\delta f(\r) ,  \label{fetot1}  \\ 
&& f_{\bar{\Delta}_d}(\r) = - {|\bar{\Delta}_d(\r)|^2\over V_d}  
- T\sum_n \int { d^2\k\over (2\pi)^2} 
{\rm ln} \left ( {\widetilde{W}_{n\k}(\r) 
\over W_{n\k} } \right )  ,  
\label{feH0}  \\
&& f_s(\r)= \Delta^2_0(T) N_0 \{ L_s'(T,\v_s) \Dd^{\prime}_s(\r) + 
L_s''(T,\v_s) \Dd^{\prime\prime}_s(\r)+[c_s +\eta_s'(T,\v_s)] 
[\Dd^{\prime}_s(\r)]^2
+[c_s +\eta_s''(T,\v_s)] [\Dd^{\prime\prime}_s(\r)]^2
 \},      
 \label{fetots}  \\
&& f_{xy}(\r)=F_4(\r)+F^{OZ}(\r)+\Delta^2_0(T) N_0 \left \{
[c_{xy}+\eta_{xy}'(T,\v_s)] [\Dd^{\prime}_{xy}(\r)]^2
+[c_{xy} +\eta_{xy}''(T,\v_s)] [\Dd^{\prime\prime}_{xy}(\r)]^2 
 \right \}, 
\label{fetotdxy} \\
&& \;\;\; F^{OZ}(\r) =  \Delta^2_0(T)N_0 \;Q^{OZ}(T)\,
{e\over mc} B(\r)\bar{\D}_{xy}^{\prime\prime}(\r), \label{feOZ} \\
&& \;\;\; F_4(\r)= \Delta^2_0(T) N_0 \left \{ L_{xy}'(T,\v_s) \Dd^{\prime}_{xy}(\r) + 
L_{xy}''(T,\v_s) \Dd^{\prime\prime}_{xy}(\r) \right \}, \label{fe4} 
\end{eqnarray}
where $'$ and $''$ indicate real and imaginary parts, respectively, 
$\Dd_i=\D_i/\Delta_0(T)$, $i=s,d_{xy}$, $c_s=(-V_sN_0)^{-1}-2c_d$, 
$c_{xy}=(-V_{xy}N_0)^{-1}-c_d$, with $c_d=(-V_dN_0)^{-1}$, $B(\r)$ is the 
magnetic induction, and $\widetilde{W}_{n\k}(\r)$, $Q^{OZ}(T)$, $L_i(T,\v_s)$ 
and $\eta_i(T,\v_s)$ are defined in Appendix A. $f_{\bar{\Delta}_d}$ in Eq. 
(\ref{feH0}) is the free energy density in association with the dominant 
$d_{x^2-y^2}$ component in the absence of $\D_s$ and $\D_{xy}$. $f_s$ and 
$f_{xy}$ are the free energy densities for the $s$ and $d_{xy}$ components,
respectively. $\delta f$ involves terms of high orders in $v_s$, $\D_i$, 
derivatives of $v_s$, and mixed $s$ and $d_{xy}$ terms. Note in Eqs. (\ref{fetots}) 
and (\ref{fetotdxy}) the quadratic terms in $D_i$, $i=s,d_{xy}$ can be easily 
reformulated to coincide with the familiar GL form: $(c_i+{\eta^\prime+
\eta^{\prime\prime} \over 2}) |\bar{\D}_i(\r)|^2+{\eta^\prime-\eta^{\prime\prime} 
\over 4} [\bar{\D}_i(\r)^2+\bar{\D}^*_i(\r)^2]$. \cite{footnote4} In addition, 
derivatives of $\D_i$ terms are absorbed into the powers of $\D_i$ terms by 
partial integration, as indicated in Appendix A.

With the free energy in Eqs. (\ref{fetot1})-(\ref{fe4}) we are in a position 
to investigate the vortex state. We will first show general results for an 
arbitrary $\v_s(\r)$ distribution \cite{note1} in Sec. III, and apply the theory 
to the single vortex case in Sec. IV.

\section{Order parameters}

\subsection{$d$-wave pairing order parameter $\bar{\Delta}_d(\r)=
\Delta_0+\D_d(\r) $}

In studying the dominant $d_{x^2-y^2}$ component, $\D_s$ and $\D_{xy}$ can be 
set to zero since the $\bar{\Delta}_d$-$\D_s$ and $\bar{\Delta}_d$-$\D_{xy}$  
mixing terms appear in higher orders of $v_s$ or its derivatives. The gap 
equation $\partial f_{\bar{\Delta}_d} / \partial \bar{\Delta}_d(\r) =0$ produces
\begin{eqnarray}
-{\bar{\Delta}_d(\r)\over V_d} = T\sum_n \int { d^2\k\over (2\pi)^2} 
{ \bar{\Delta}_d(\r) \cos^2 2\varphi \over \widetilde{W}_{n\k}(\r) },
 \label{dgap}
\end{eqnarray} 
where $\widetilde{W}_{n\k}(\r)$ is defined in Eq. (\ref{Wtilde}) in Appendix A.
At $H=0$, Eq. (\ref{dgap}) reduces to 
\begin{eqnarray}
1 = -V_d T\sum_n \int{ d^2\k\over (2\pi)^2} {\cos^2 2\varphi \over W_{n\k}} 
= -V_d \int { d^2\k\over (2\pi)^2} {\cos^2 2\varphi~ {\rm tanh}(E_\k/2T)
\over 2 E_\k}, 
\label{dgapH0}
\end{eqnarray} 
which yields the well-known $T_c$ formula $c_d =(1/2) {\rm ln} (2e^\gamma
\omega_D/\pi T_c)$, with $\omega_D$ the BCS cutoff and $\gamma\simeq 0.5772$ 
the Euler constant, as well as the asymptotic behaviors of the gap maximum
\begin{equation}
 \Delta_0(T)\simeq \left\{
 \begin{array}{ll}
 3.54 [T_c(T-T_c)]^{1/2}, \;\;\;\;\; & T\gg \Delta_0(T)\\
 2.14 T_c - 0.39 (T/T_c)^3 T_c, \;\;\;\; &  T\ll \Delta_0(T)
  \end{array}
 \right. .    \label{dgapH0asymp}
\end{equation}

The presence of vortices depletes the pairing order parameter $\Delta_0$ by 
$\D_d(\r)$. Sufficiently far from the vortex cores, either $T$ or $\Delta_0(T)$ 
is larger than $v_s(\r)k_F$ in the whole $T$ regime, and thus we can treat 
$v_s(\r)k_F$ and $\D_d(\r)$ perturbatively. It is straightforward to show that 
the normalized real part of $\D_d(\r)$ is 
\begin{eqnarray}
&& {\D^\prime_d(\r)\over \Delta_0(T)} \simeq 
\left \{  \begin{array}{ll} 
- \, {1\over 3/4+ \epsilon_F \, m v^2_s(\r)/\Delta^2_0(T)} 
{\epsilon_F \, mv^2_s(\r) \over \Delta^2_0(T)} ,  &  \;\;\;\; {\rm GL} \; 
{\rm regime} \\
- \,  {2 \;({\rm ln}\,2) \over 1- [9\zeta(3)T^3
- 4 \,({\rm ln}\,2) \epsilon_F T m v^2_s(\r)]/\Delta^3_0(T)} 
{\epsilon_F\, mv^2_s(\r)\over \Delta^2_0(T)} {T\over\Delta_0(T)} , & \;\;\;\;
v_sk_F \ll T\ll\Delta_0(T) \\
- \,{  \sum_{l=\pm 1} |\cos (\theta+l {\pi\over 4} ) |^3  \over 
3- [\sum_{l=\pm 1} |\cos (\theta+l {\pi\over 4} ) |^3] [v_s(\r)k_F]^3/
\Delta^3_0(T)} {\epsilon_F\, mv^2_s(\r)\over \Delta^2_0(T)} 
{v_sk_F\over\Delta_0(T)} , &   \;\;\;\;  T\ll v_sk_F\ll \Delta_0(T) 
\end{array}  \right.  ,
\label{dOP}
\end{eqnarray}
where $\epsilon_F$ is the Fermi energy and $\zeta(3)$ is the Riemann function. 
The imaginary part of $\D_d(\r)$ driven by the derivatives of the supercurrent 
is unimportant compared with $\D^{\prime}_d(\r)$ in the whole temperature region, 
for its driving term $F^{(1)}_d$ in Eq. (\ref{feder1}) is very small as discussed 
in Appendix A.

\subsection{Field-Induced $\D_s(\r)$} 

From Eqs. (\ref{fetot1}) and (\ref{fetots}) we obtain the gap equation for 
the $s$ component, $\partial f_s(\r) / \partial \D_s(\r) =0$, which gives
\begin{eqnarray}
&&\Dd^{\prime}_s(\r) \simeq -{L^\prime_s(T,\v_s) \over 2 
[c_s+\eta^\prime_s(T,\v_s)]} ,  \;\;\;\;\;\;\;\;\; 
\Dd^{\prime\prime}_s(\r) \simeq -{L^{\prime\prime}_s(T,\v_s) \over 2 
[c_s+\eta^{\prime\prime}_s(T,\v_s)]},   \label{sgap} 
\end{eqnarray}
where $L^\prime_s(T,\v_s)$, $L^{\prime\prime}_s(T,\v_s)$, $\eta^\prime_s(T,\v_s)$
and $\eta^{\prime\prime}_s(T,\v_s)$ are defined in Appendix A.

For $T>\v_s\cdot \k_F$, we can plug $L_s$ and $\eta_s$ obtained in
Eqs. (\ref{Qsr1})-(\ref{Qsi2}) into Eq. (\ref{sgap}) to get the scaling 
functions of $\Dd^{\prime}_s(\r)$ and $\Dd^{\prime\prime}_s(\r)$,
\begin{eqnarray}
\Dd^{\prime}_s(\r) &\simeq &  G_{s1}\left ({\Delta_0(T)\over T},
{\epsilon_Fmv^2_{s}(\r) \over T^2}\right ) {\epsilon_Fm [v^2_{sx}(\r)- 
v^2_{sy}(\r)] \over T^2}, \label{sr}\\
\Dd^{\prime\prime}_s(\r)&\simeq & G_{s2} \left ({\Delta_0(T)\over T},
{\epsilon_Fmv^2_{s}(\r) \over T^2} \right ) {\epsilon_F\partial_xv_{sx}(\r) 
\over T^2},    \label{si}
\end{eqnarray}
where 
\begin{eqnarray}
G_{s1}(d,z)&=&-{h_{1s}(d)\over c_s-h_{3s}(d)+2 d^2 h_{2s}(d)+z h_{4s}(d)}, 
\label{fs1} \\
G_{s2}(d,z)&=&-{2h_{2s}(d)\over c_s-h_{3s}(d)+z h_{4s}(d)}, \label{fs2} 
\end{eqnarray}
with $h_{1s}$, $\cdots$, $h_{4s}$ defined in Eqs. (\ref{h12i}) and (\ref{h34i}). 

It is easy to see that in the GL regime,
\begin{eqnarray}
&&\Dd^{\prime \,{\rm GL} }_s(\r)\simeq - \, {1 \over c_s+{3\over 4}{\Delta^2_0(T)
\over \pi^2 T^2} + 2{\epsilon_F m v^2_s(\r)\over \pi^2 T^2} }  
{\epsilon_Fm[v^2_{sx}(\r)-v^2_{sy}(\r)] \over \pi^2 T^2},  \label{srGL} \\
&&\Dd^{\prime\prime \,{\rm GL} }_s(\r)\simeq - {1 \over c_s- 
{\Delta^2_0(T)\over 4\pi^2T^2}+2{\epsilon_F m v^2_s(\r)\over \pi^2
T^2}  } \;{\epsilon_F\partial_x v_{sx}(\r)  \over \pi^2 T^2}.
\label{siGL} 
\end{eqnarray}
It follows that the ratio between the real and imaginary parts of $\D_s$ 
\begin{eqnarray}
{\Dd^{\prime \,{\rm GL} }_s(\r)\over \Dd^{\prime\prime \,{\rm GL} }_s (\r)}\simeq
 { m [v^2_{sx}(\r)-v^2_{sy}(\r)] \over \partial_x v_{sx}(\r)} 
 \label{sratioGL}
\end{eqnarray}
is of order unity in most of the region outside the core where $v_s(\r)\sim 1/r_0$ 
and $\partial_x v_{sx}(\r)\sim 1/r^2_0$ with $r_0$ the distance to the closest 
vortex core center.

In the case of $\v_s\cdot \k_F <T \ll \Delta_0(T)$,
\begin{eqnarray}
&&\Dd^{\prime \, {\rm low T}}_s(\r)\simeq - \, {({\rm ln}\,2) \over c_s+{1\over 2}
+ {\epsilon_F mv^2_s(\r)\over 4T\Delta_0 }  } \, {T\over \Delta_0}\,
{\epsilon_F m [v^2_{sx}(\r)-v^2_{sy}(\r)]\over \Delta^2_0} ,  \label{srlowT} \\
&&\Dd^{\prime\prime \, {\rm low T}}_s(\r) \simeq - {1-2\, ({\rm ln}\,2)  
{T\over \Delta_0} \over  c_s-{1\over 2} +2 ({\rm ln}\,2) {T\over \Delta_0}
+ {\epsilon_F  m v^2_s(T)\over 4T\Delta_0} }\, {\epsilon_F \, \partial_x 
v_{sx}(\r)\over \Delta^2_0}.  \label{silowT}
 \end{eqnarray}
So the ratio between the real and imaginary parts becomes
\begin{eqnarray}
{\Dd^{\prime \, {\rm low T}}_s(\r)\over \Dd^{\prime\prime \, 
{\rm low T}}_s(\r)} \simeq ({\rm ln}\,2) 
{T\over \Delta_0} \, { m [v^2_{sx}(\r)-v^2_{sy}(\r)] \over \partial_x v_{sx}(\r)} . 
\label{sratiolowT}  
\end{eqnarray}
Comparing Eqs. (\ref{sratioGL}) and (\ref{sratiolowT}) we find an extra prefactor 
$T/\Delta_0$ is acquired at low $T$, indicative of a suppressed real part of the 
$s$ component with decreasing $T$. This is an interesting observation in the 
present work, the consequence of which on the structure of a single vortex 
will be discussed in Sec. IV. 

For $T<\v_s\cdot\k_F$, which can be achieved either by lowering $T$ in a 
certain spatial position or approaching the core region at a certain $T$, 
the nonlinear effects dominate over the thermal effects and the prefactor 
$T/\Delta_0$ in Eq. (\ref{sratiolowT}) is expected to be replaced by 
$v_sk_F/\Delta_0$. So the real part is negligibly small, and the imaginary 
part at $T=0$ is 
\begin{eqnarray}
&&\Dd^{\prime\prime \, {T=0}}_s(\r)\simeq - {1-S_\theta \,v_s(\r)k_F/
\Delta_0 \over c_s -1/2+S_\theta \,v_s(\r)k_F/2\Delta_0 }
{\epsilon_F \, \partial_x v_{sx}(\r) \over \Delta^2_0},   \label{siT0}
 \end{eqnarray}
where $S_\theta=\sum_{l=\pm 1} \left |\cos (\theta+l {\pi\over 4}) \right |$.
It is interesting to find from Eqs. (\ref{siT0}) and (\ref{silowT}) that the 
factor $T/\Delta_0$ at $\v_s\cdot\k_F<T$ is replaced by $v_sk_F/\Delta_0$ 
(up to some prefactor) at $T<\v_s\cdot\k_F$, reflecting the nonlinear effect 
due to the Doppler energy shift. Its effect on the free energy and the 
penetration depth in the Meissner state of a $d$-wave superconductor was 
extensively discussed in the previous papers by the present authors. \cite{LHWprb}

\subsection{Field-Induced $\D_{xy}(\r)$} 

\begin{figure}[h]
\begin{picture}(180,150)
\leavevmode\centering\includegraphics{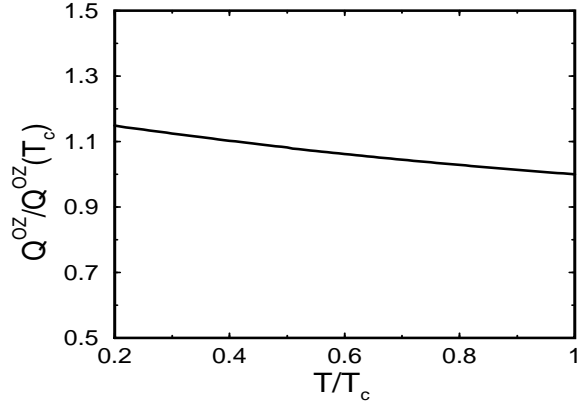}
\end{picture}
\caption{Normalized $Q^{OZ}$ as a functions 
of $T/T_c$. $c_d=3$ is chosen.}
\label{qozt}
\end{figure}

As shown in Eq. (\ref{fetotdxy}), there are two terms, $F^{OZ}$ and $F_4$,
competing in driving the induced $\dxy$ component. The first term $F^{OZ}$, 
the so-called {\it orbital Zeeman} term,\cite{Koyama} can be rewritten as
\begin{equation}
F^{OZ} (\r) = - {\bf M}(\r)\cdot {\bf B}(\r),\label{OZ1}
\end{equation} 
with ${\bf M}(\r)$ the effective magnetic moment associated with the internal 
orbital current of Cooper pairs in a $d_{x^2-y^2}+id_{xy}$-wave superconductor. 
A general derivation of this spontaneous magnetization is presented in Appendix B. 
As also discussed there, ${\bf M}(\r)$ is proportional to the particle-hole 
asymmetry $\alpha$ (see Eq. (\ref{DOS})),  making the Zeeman orbital effects 
very small for general density of states. It is therefore worth pointing out 
that analyses of the $\dx2-y2$-$\dxy$ mixing problem based solely on a 
symmetry-based analysis of the orbital moment \cite{Balatsky} may lead to 
unrealistic results. Besides, it is important to note the very weak temperature 
dependence of the coefficient $Q^{OZ}$ of $F^{OZ}$, as shown in Fig. \ref{qozt}.

The second term, $F_4$, is in the order ${\cal O}(v_s^4)$ and contains driving 
terms for $\D^{\prime\prime}_{xy}$ as well as $\D^{\prime}_{xy}$. This term can 
be significant particularly if the system has small or zero particle-hole 
asymmetry. Note that in the $s$-wave case, it is irrelevant except for very 
short length scales of order the core size, due to the nonvanishing leading 
(${\cal O}(v_s^2)$) term.

\begin{figure}[h]
\begin{picture}(180,170)
\leavevmode\centering\includegraphics{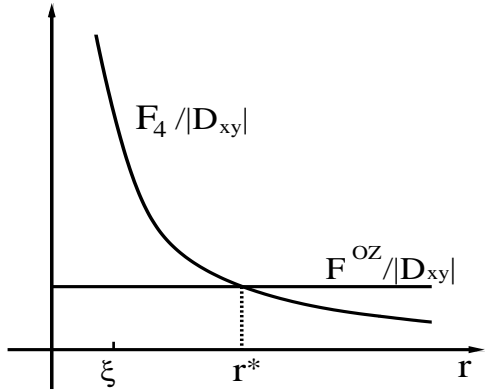}
\end{picture}
\caption{Schematic comparison of $F^{OZ}$ and $F_4$.}
\label{dxydriving}
\end{figure}

The orbital Zeeman term has been invoked by Laughlin \cite{Laughlin}
and Balatsky \cite{Balatsky} as the effect driving a putative transition 
to a time reversal symmetry-breaking $d_{x^2-y^2}+id_{xy}$ state induced 
by field. It is therefore particularly interesting to investigate the 
relative importance of $F^{OZ}$ and $F_4$ within the current approach. 
For fields $H\gg H_{c1}$, the overlap of vortices leads to nearly homogeneous 
$B(\r) \simeq H$ in space and hence $F^{OZ}/|\D_{xy}|$ can be taken as a 
constant over the bulk. On the other hand, $F_4/|\D_{xy}|$, which scales 
as $(\xi/r)^4$, is strongly space dependent. It increases rapidly when 
approaching the vortex core, but decays into the bulk. Thus we expect a 
critical radius $r^*$ beyond which $F^{OZ}$ dominates over $F_4$, but for 
$r<r^*$ $F_4$ becomes more important and determines the structure. $r^*$ 
can be estimated from $F_4(r^*)/F^{OZ}(r^*) \sim 1$. It turns out to  
grow with increasing $H$ and decreasing $T$. In the GL and low $T$ regime, 
we have
\begin{equation}
r^{*}(T,H) = \left \{ \begin{array}{ll}
\xi \left ({\epsilon_F\over eH/mc} \right )^{1/4} 
\sim \xi \left ({R_H\over \lambda_F}\right )^{1/2},  \;\;\;\;\; & GL \\
\xi \left ({\Delta_0(T)\over T}\right )^{1/4}\left ({\epsilon_F\over eH/mc}
\right )^{1/4} \sim \xi \left ({\Delta_0(T)\over T}\right )^{1/4}\left (
{R_H\over\lambda_F} \right )^{1/2},  \;\;\;\;\; & {\rm low} \; T
\end{array}  \right. ,
\label{r2}
\end{equation}
with $\lambda_F$ the Fermi wavelength, $\xi=v_F/\pi\Delta_0$ the superconducting 
coherence length, and $R_H=\sqrt{c/eH}$ the average intervortex distance. We 
summarize the above estimates in Table I, 
and display schematically the competition between the two driving terms in 
Fig. \ref{dxydriving}. Note the ``low T" results for $F_4$ obtain only down to
temperatures above the local superfluid velocity Doppler shift $v_sk_F$.
For lower $T$, the perturbation calculation for $F_4$ that we present in 
Appendix A.3 breaks down because of the zero modes in $E_\k$. Instead one has to  
get a full expression for the local Doppler shift and the derivatives of the 
supercurrents which is suitable for perturbation expansion {\it after} 
integrating over momenta. This is apparently too complicated to be achieved 
analytically and thus requires a self-consistent numerical work, which is beyond 
the scope of the present paper. However, based on the  nonlinear results for the 
$s$ component and $Q^{OZ}$, we do not expect anything qualitatively new in the 
extremely low $T$ case compared with the $\v_s\cdot \k_F\ll T\ll\Delta_0(T)$ case.

\begin{table}
\caption{ Order of magnitude of terms driving $\dxy$ in free energy {\it 
outside of core region}  }
\label{table2}
\begin{tabular}{ccc}
Term   &   GL regime $(\Delta_0 \ll T)$   &   Low T $(v_s k_F\ll T\ll \Delta_0)$ 
\\   \hline\hline
$F^{OZ}$  & ${N_0\over \epsilon_F}{eH\over mc} \Delta_0 {\cal D}_{xy}$
& ${N_0\over \epsilon_F}{eH\over mc}\Delta_0 {\cal D}_{xy}$  \\ \hline
$F_4$   &  $N_0\left({\xi\over r} \right)^4 \Delta_0{\cal D}_{xy}$ &
$N_0\left({\xi\over r} \right)^4 \left({\Delta_0\over T}\right) 
\Delta_0{\cal D}_{xy}$ 
\end{tabular}
\end{table}

The resulting $d_{xy}$-wave order parameter, for $\xi<r<r^*$ where $F_4$ 
dominates, is  
\begin{eqnarray}
&&\Dd^{\prime}_{xy}(\r) \simeq -{L^\prime_{xy}(T,\v_s) \over 2 
[c_{xy}+\eta^\prime_{xy}(T,\v_s)]} ,  \;\;\;\;\;\;\;\;\; 
\Dd^{\prime\prime}_{xy}(\r) \simeq -{L^{\prime\prime}_{xy}(T,\v_s) \over 2 
[c_{xy}+\eta^{\prime\prime}_{xy}(T,\v_s)]} ,  \label{dxy1} 
\end{eqnarray}
where $L^\prime_{xy}(T,\v_s)$, $L^{\prime\prime}_{xy}(T,\v_s)$, $\eta^\prime_{xy}
(T,\v_s)$, and $\eta^{\prime\prime}_{xy}(T,\v_s)$ are defined in Eqs. (\ref{lrxy}), 
(\ref{lixy}), (\ref{Qdxyr2}) and (\ref{Qdxyi2}), and their asmptotic behaviors 
can be found in Table II. In the region $r^*<r<R_H$ where $F^{OZ}$ dominates, 
the $d_{xy}$ component reads
\begin{eqnarray}
&& \Dd^{\prime}_{xy}(\r) \simeq 0, \;\;\;\;\;\;\;\;\;\;\;\;\;\;\;\;
\Dd^{\prime\prime}_{xy}(\r) \simeq - {Q^{OZ}(T) \over 2 [c_{xy}+
\eta^{\prime\prime}_{xy}(T,\v_s)]} {eH\over mc}, \label{dxy2} 
\end{eqnarray}
where $Q^{OZ}(T)$ defined in Eq. (\ref{qoz}). Eq. (\ref{dxy2}) and the asymptotic
behaviors of $Q^{OZ}$, $\eta^\prime_{xy}(T,\v_s)$, and $\eta^{\prime\prime}_{xy}
(T,\v_s)$ shown in Tabel II lead to 
\begin{eqnarray}
\Dd^{\prime\prime}_{xy}(\r) \simeq \left \{
\begin{array}{ll}
{1\over 2}\alpha(2c_d-1) [c_{xy}-{3\over 8}{\Delta_0^2(T)\over \pi^2T^2} + 
{\epsilon_Fmv^2_s(\r)\over \pi^2T^2}]^{-1} {1 \over \epsilon_F}  {eH\over mc}
\;\;\; & \Delta_0(T)\ll T \\
{1\over 2}\alpha [{\rm ln}({4\omega_D\over \Delta_0})-{1\over 2}-
6({\rm ln}2){T\over \Delta_0}] [c_{xy}-{1\over 2}+2({\rm ln} 2){T\over \Delta_0} 
+ {\epsilon_Fmv^2_s(\r)\over 4\Delta_0 T}]^{-1} {1 \over \epsilon_F} {eH\over mc} 
\;\;\; & T\ll \Delta_0(T) 
\end{array}
\right. .   \label{outerring}
\end{eqnarray}

It is interesting to estimate the value of $r^*$ in a real material. It follows 
from Eq. (\ref{r2}) that for $T\sim T_c$, $r^{*GL}/R_H\simeq \sqrt{\xi/\lambda_F} 
(H/H_{c2})^{1/4} $. For high-$T_c$ cuprates, $\xi/\lambda_F\sim 10$, and thus 
$r^{*GL}>R_H$ for fields $H> 0.01H_{c2}$. In materials with larger $\xi/\lambda_F$, 
$r^{*GL}$ becomes order of $R_H$ for smaller fields. Since from Eq. (\ref{r2}) 
$r^*$ increases with decreasing temperatures, it seems unlikely that the orbital 
Zeeman free energy plays an important role in determining the local order parameter
in the vortex state for fields in the Tesla range.

\begin{figure}[h]
\begin{picture}(180,160)
\leavevmode\centering\includegraphics{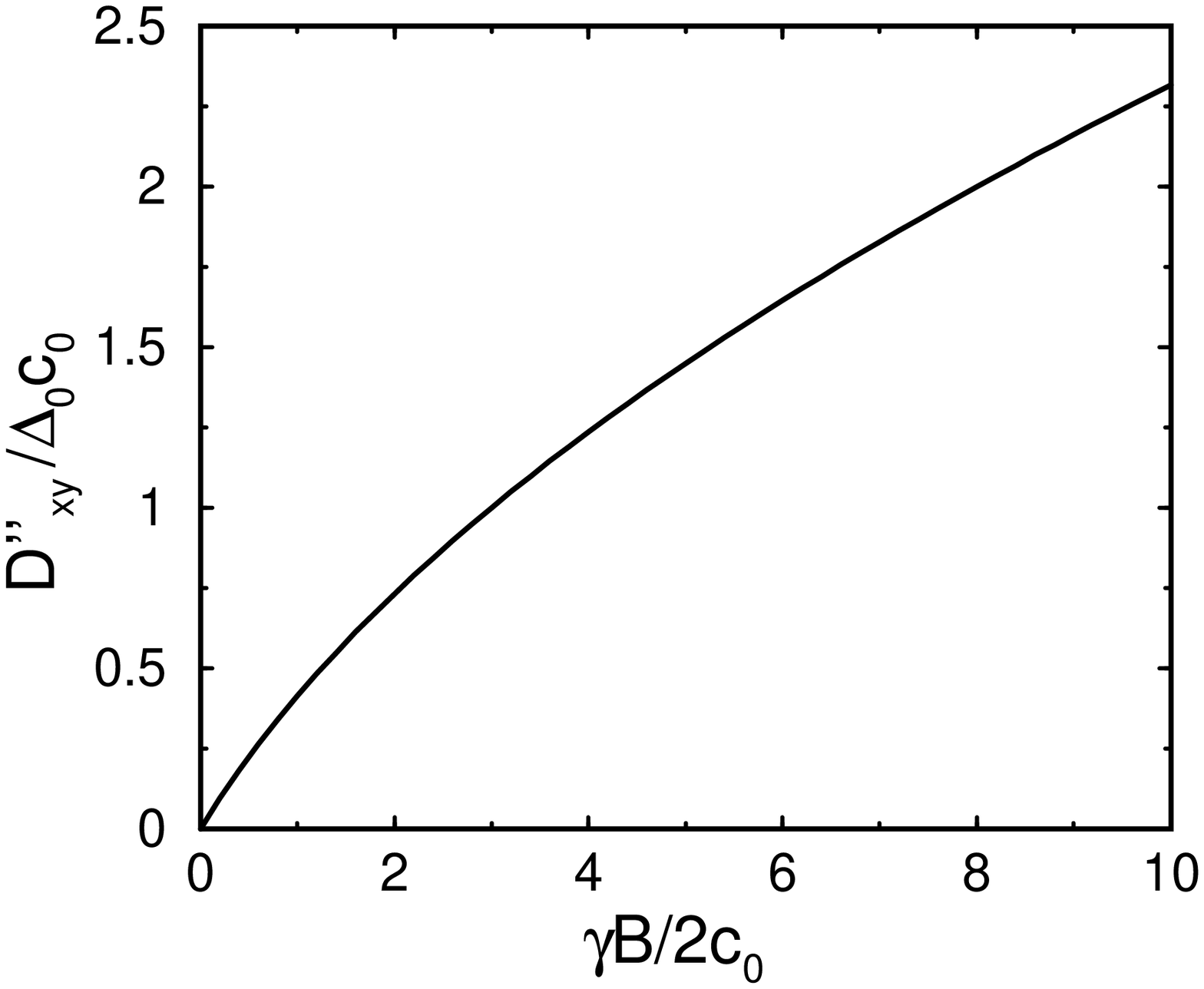}
\end{picture}
\caption{Magnetic field dependence of $\bar{\D}^{\prime\prime}_{xy}$
for $r>r^*$ at low $T$.}
\label{bdependence}
\end{figure}

We would like to make some further remarks on the magnetic field dependence of 
$\bar{\D}^{\prime\prime}_{xy}$ for $r>r^*$ at low temperatures. 
$\bar{\D}^{\prime\prime}_{xy}$ shown in Eqs. (\ref{dxy2}) and (\ref{outerring}) 
are obtained from minimizing the free energy density $f_{xy}(\r)$ in Eq. 
(\ref{fetotdxy}). This free energy density is up to quadratic in 
$\bar{\D}^{\prime\prime}_{xy}$, which is sufficient for low field and for generic 
$c_{xy}\gg 1$. In the case of larger field and/or special case of $c_{xy}$ close 
to or smaller than 1, one has to include the free energy density term cubed in 
$\bar{\D}^{\prime\prime}_{xy}$. This term can be easily found to be 
$(\bar{\D}^{\prime\prime}_{xy})^3/3$ with the spatial dependence of the
coefficient neglected. Thus the free energy density up to cubed in 
$\bar{\D}^{\prime\prime}_{xy}$ for $r>r^*$ at low temperatures can be written as
\begin{eqnarray}
\tilde{f}_{xy}= -\gamma B\bar{\D}^{\prime\prime}_{xy}+c_0
(\bar{\D}^{\prime\prime}_{xy})^2+ {1\over 3}(\bar{\D}^{\prime\prime}_{xy})^3,
\label{feout}
\end{eqnarray}
where $c_0 = c_{xy}+\eta^{\prime\prime}_{xy}(T,\v_s)$ and $\gamma = \Delta^2_0 
N_0 Q^{OZ} e/mc$. Minimizing $\tilde{f}_{xy}$ with respect to 
$\bar{\D}^{\prime\prime}_{xy}$ we immediately get
\begin{eqnarray}
\bar{\D}^{\prime\prime}_{xy} \simeq \sqrt{c_0^2+\gamma B/2}-c_0 .
\label{douter}
\end{eqnarray}
It is obvious that there is crossover of the linear-$B$ dependence of 
$\bar{\D}^{\prime\prime}_{xy}$ for $\gamma B/2\ll c^2_0$ to square root of $B$ 
dependence of $\bar{\D}^{\prime\prime}_{xy}$ for $\gamma B/2\gg c^2_0$.
This interesting behavior is shown in Fig. (\ref{bdependence}).
 
In a periodic vortex lattice, the supercurrent field $\v_s(\r)$ in the London 
approximation reads
\begin{eqnarray}
\v_s(\r)={\pi\over m} \sum_{{\bf K}\in G} {i{\vec \nabla}_\r
(e^{-i {\bf K}\cdot \r})\over K^2+\lambda^{-2}} ,
\end{eqnarray} 
and thus there will be special symmetry points where $\v_s=0$. At these points, 
a careful examination shows that the coefficients $L^{\prime}_{xy}$ and 
$L^{\prime\prime}_{xy}$ in $F_4$ (Eq. (\ref{fe4})) vanish, and that $\D_{xy}$ 
is driven entirely by $F^{OZ}$, as shown in Eq. (\ref{douter}). This coincides 
with the numerical result of Yasui and Kita. \cite{Kita}

\begin{table}
\caption{Asymptotic behaviors of $L^\prime_{xy}$, $L^{\prime\prime}_{xy}$, 
$Q^{OZ}$, $\eta^\prime_{xy}$, and $\eta^{\prime\prime}_{xy}$ (See Appendix A 
for definitions of $\v_s$-dependent coefficients $U$)}
\label{table1}
\begin{tabular}{ccc}
Term   &   GL regime $(\Delta_0 \ll T)$   &   Low T $(v_s k_F\ll T\ll \Delta_0)$ \\
\hline 
$L^\prime_{xy}$  &  $({1\over 2} U^{\prime}_1+{3\over 2} U^{\prime}_2-{1\over 2} 
U^{\prime}_3)/(\pi T)^4$  &  $({1\over 12} U^{\prime}_1+{1\over 2} U^{\prime}_2
+ {1\over 24} U^{\prime}_3)/\Delta_0^3 T$ \\ \hline
$L^{\prime\prime}_{xy}$  &  $(-{3\over 2} U^{\prime\prime}_1+{1\over 16} 
U^{\prime\prime}_2)/(\pi T)^4$  &  $(-{1\over 4} U^{\prime\prime}_1
+ {1\over 24} U^{\prime\prime}_2)/\Delta_0^3 T$ \\ \hline
$Q^{OZ}$  &   $-(\alpha/\epsilon_F)(2 c_d-1)$  &  
$-(\alpha/\epsilon_F)[{\rm ln}(4\omega_D/\Delta_0)-1/2-
6({\rm ln}2)(T/\Delta_0)]$   \\ \hline
$\eta^{\prime}_{xy}$ & $ -(1/8)\Delta_0^2(T)/\pi^2 T^2+ 
\epsilon_Fmv^2_s(\r)/\pi^2T^2 $  & $ \epsilon_Fmv^2_s(\r)/4\Delta_0 T $ \\ \hline
$\eta^{\prime\prime}_{xy}$  &  $-(3/8) \Delta_0^2(T)/\pi^2T^2 + 
\epsilon_Fmv^2_s(\r)/\pi^2T^2 $  &  $-1/2+2({\rm ln} 2)T/\Delta_0 +
\epsilon_Fmv^2_s(\r)/4\Delta_0 T  $  
\end{tabular}
\end{table}

Up to now, we have not included the effect of the $d_{x^2-y^2}$ order-parameter 
suppression $\D_d$ on the $d_{xy}$ component, since we have shown in Sec. III A 
that $\D_d$ is negligibly small in the bulk region, in which we are primarily 
interested. However, the mechanism for a transition to a $ d_{x^2-y^2}+id_{xy}$
state proposed by Ramakrishnan\cite{Ramakrishnan} involves precisely the interplay 
between $\D_d$ and the  supercurrent near the core, leading potentially to local
$d_{x^2-y^2}+id_{xy}$ patches  which then overlap at some critical field. Motivated 
by this suggestion, we examine the free energy terms including this effect within 
our approximation. These terms, which  we refer to as the order parameter
suppression terms $F^{OPS}$, are obtained from Eq. (A9-A12) (and have been already
neglected in arriving at Eq. (20)). It can easily be seen that the $\D_{xy}$ 
component couples to derivatives of either $\D_d(\r)$ or of $\v_s$ in these terms, 
indicative of a pure nonlocal effect as found by Ramakrishnan. However, for 
$\v_s\cdot\k_F<T$, our surprising finding is that up to leading order, terms
including derivatives of $\D_d(\r)$ vanish, leaving 
\begin{eqnarray}
&& F^{OPS} = -\int d^2 r \left (T\sum_n \int {d^2\k\over (2\pi)^2} 
{ \epsilon_\k \cos^2 2\varphi \over W^2_{n\k}} \right )
\D_d(\r) [\nabla_\r\times \v_s(\r)]_z \D^{\prime\prime}_{xy}(\r). \label{feOPS}
\end{eqnarray}
Since the $[\nabla_\r\times \v_s(\r)]_z$ factor is dominated by the vector 
potential $\A(\r)$ rather than $\nabla \phi$ part in $\v_s(\r)$, it may be 
replaced by $eB(\r)/(mc)$. Comparing Eqs. (\ref{feOPS}) and (\ref{feOZ}), we see 
that $F^{OPS}$ in the core region is of the same order as $F^{OZ}$ in the
bulk (in particular, it is also proportional to the particle-hole asymmetry 
of the normal state), and may be viewed as the leading correction to $F^{OZ}$, 
if $\Delta_0(T)$ in $F^{OZ}$ is replaced by $\bar{\Delta}_d(\r)=\Delta_0(T) +
\D_d(\r)$. It is therefore clear from our previous discussion of $F^{OZ}$
that $F^{OPS}$ also gives in fact a very small effect even near the vortex core.

\section{Single vortex}

The results obtained in Sec. III enable us to compare the subdominant order 
parameters, at various temperatures, in the presence of vortices characterized 
by a certain superfluid velocity field $\vs (\r)$. \cite{note1} Our purpose 
in this section is to test these results in the concrete case of a single 
isolated vortex.

We use cylindrical coordinates $\r=(r,\theta)$  with the origin located at 
the vortex core center.  The phase field of the order parameter $\phi(\r)=\theta$ 
leading to $\nabla \phi(\r)=-(1/2mr)\hat{\theta}$. In the spatial regime of 
interest, the magnetic field is roughly homogeneous and thus ${\bf A}(\r)
= Br\hat{\theta}$. For $r\ll R_H$, $|\nabla \phi(\r)/2m|\gg eA(\r)/(mc)$, so 
we can neglect $\bf A$, and simply write the supercurrent components and their 
derivatives as
\begin{equation}
v_{sx}\simeq {\sin\theta\over 2mr}\, , \;\;\;\;\;\;\;\;\;\; 
v_{sy}\simeq -{\cos\theta\over 2mr}\, , \;\;\;\;\;\;\;\;\;\;
\partial_x v_{sx}(\r)\simeq - {\sin 2\theta\over 2mr^2}\, ,
\label{vssingle}
\end{equation}
except for studying the orbital Zeeman term, in which case $A(\r)$ is no longer 
negligible because $\nabla\times {\bf A}$ enters.

\begin{figure}[h]
\begin{picture}(180,380)
\leavevmode\centering\includegraphics{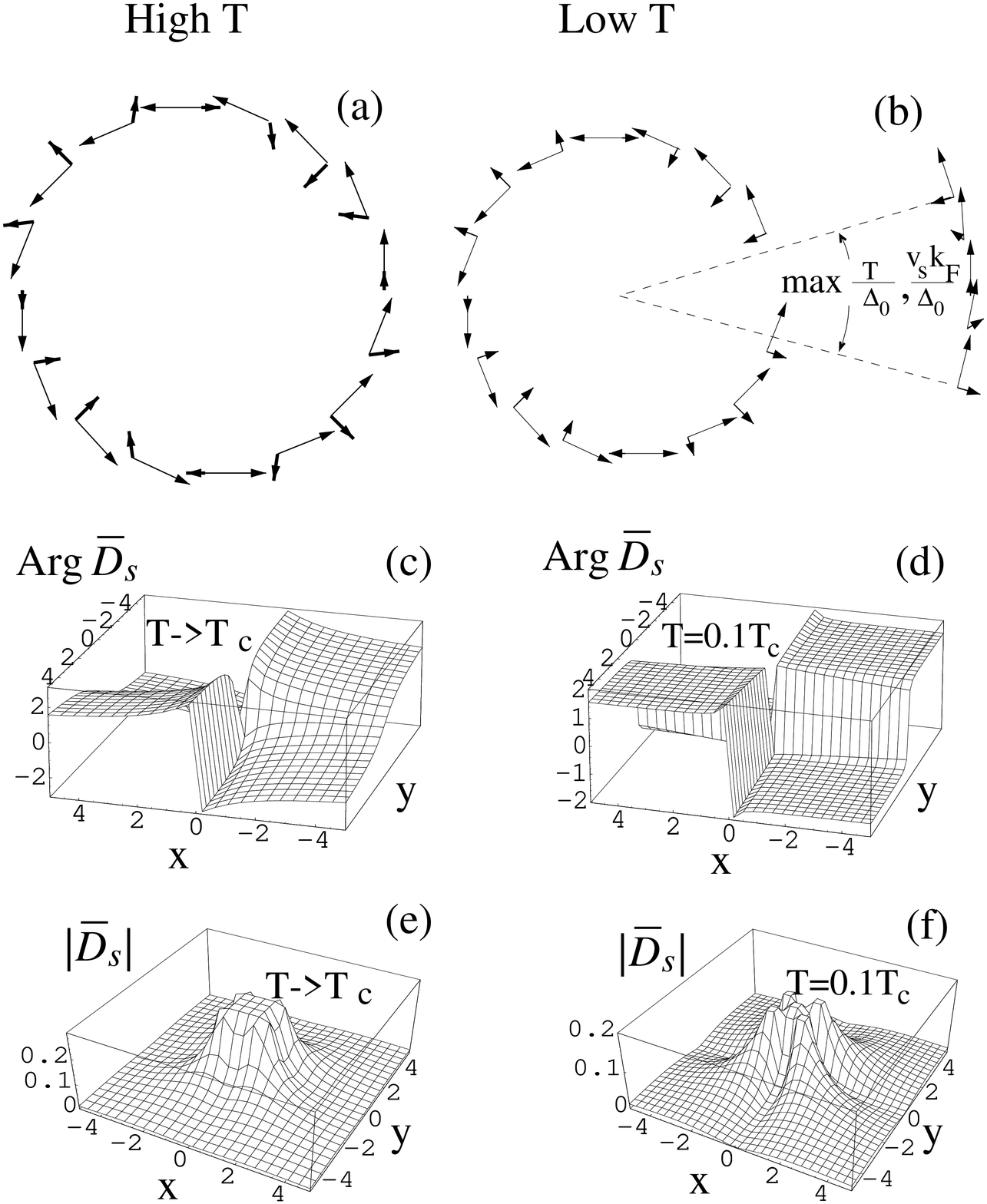}
	      \end{picture}
\caption{Single vortex structure with $d_{x^2-y^2}$ and $s$-wave 
symmetries only in the GL regime (``High T") and for $v_sk_F\ll T\ll\Delta_0$ 
(``Low T"). $c_d=3$ and $c_s=4$ are chosen throughout, and distances
are given in units of $\xi$.  a) and b) Relative phase of $s$ and $d_{x^2-y^2}$
order parameters with long arrows corresponding to $d_{x^2-y^2}$,
short to $s$, and angle between them to relative phase. c)-f) Arguments and 
magnitudes of normalized subdominant $s$ order parameter $\bar{\D}_s$.}
\label{svortexstructure}
\end{figure}

\subsection{$\dx2-y2$-$s$ mixing} 

In the GL regime, the $s$-wave subdominant order parameter has been investigated 
by many authors, \cite{RXT,Berlinsky,Franz} with substantial agreement. The 
existing results in this limit are easily shown to be recovered in the present 
theory. Inserting Eq. (\ref{vssingle}) into Eqs. (\ref{srGL}) and (\ref{siGL})
leads, in the generic case of $c_s\gg 1$, to 
\begin{equation}
\Dd^{GL}_s(\r)\simeq 0.6 c_s^{-1} {\left (\xi\over r\right )^2} (\cos 2\theta +
2 i \sin 2\theta)  = 0.3 c_s^{-1}  {\left (\xi\over r\right )^2} (3 e^{2i\theta} 
- e^{-2i\theta}) 
\end{equation}
which is consistent with the results in Ref.\cite{RXT,Berlinsky,Franz} in the 
bulk asymptotic region $r\gg \xi$. As a result, the $s$ component is of 
four-fold symmetry, and the relative winding of $s$ and $d$ components is 
uniform across the whole vortex as shown in Fig. \ref{svortexstructure}. 

\begin{figure}[h]
\begin{picture}(180,180)
\leavevmode\centering\includegraphics{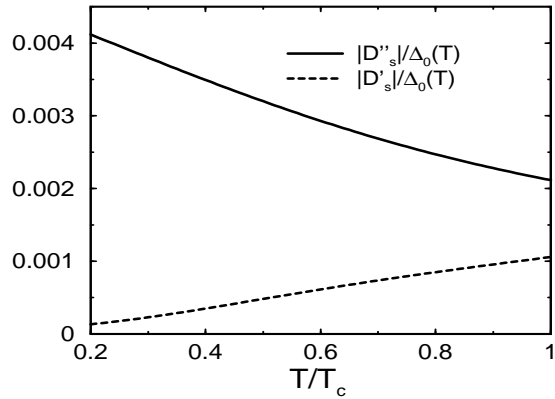}
\end{picture}
\caption{Real and imaginary parts of $\bar{\D}_s$ as functions 
of $T/T_c$ for $\v_s\cdot \k_F<T$. $r/\xi=10$, $\theta=\pi/8$, 
$c_d=3$, and $c_s=4$ are chosen.}
\label{swaveT}
\end{figure}

In the low $T$ case, as discussed in Sec. III B, $\D^{\prime}_s(\r)$ is smaller 
than $\D^{\prime\prime}_s(\r)$ by a factor of (ln 2)$T/\Delta_0$, and the vortex 
structure at low $T$ is thus expected to be qualitatively different from that 
in the GL regime. From Eqs. (\ref{vssingle}), (\ref{srlowT}), and (\ref{silowT}), 
one finds that
\begin{equation}
\Dd_s^{\rm low T}(\r)\simeq 1.23 c_s^{-1} {\left (\xi\over r\right )^2}
\left [ ({\rm ln} 2){T\over \Delta_0} \cos 2\theta +2 i \sin 2\theta \right ] .
\label{ssingle}
\end{equation}
The magnitude and phase of $\D_s^{\rm low T}$ are shown in Fig. 
\ref{svortexstructure}.  The relative winding of $s$ and $d$ components at low 
$T$ takes place in a very narrow region of real space near antinode directions 
set by max$(T/\Delta_0,v_sk_F/\Delta_0)$. In Fig. \ref{swaveT}, we show the $T$ 
dependence of $\bar{\D}^{\prime\prime}_s$ and $\bar{\D}^{\prime}_s$ in a spatial 
position in the bulk.

\subsection{$\dx2-y2$-$\dxy$ mixing}

As discussed in Sec. III C, the competition between $F^{OZ}$ and $F_4$ divides 
the outside-vortex-core region into two rings: the outer ring region $r^*<r<R_H$ 
is dominated by $F^{OZ}$ resulting in a rigid $d_{x^2-y^2}+id_{xy}$ superconducting 
state with spatially nearly constant $\D^{\prime\prime}_{xy}$ obtained in Eqs. 
(\ref{dxy2}) and (\ref{outerring}); In the inner ring $\xi<r<r^*$, $F_4$ is more 
important and a spatially varying $\D_{xy}$ is expected. Now we show the single 
vortex structure in the inner ring region. We first insert Eq. (\ref{vssingle}) 
into $U(\v_s)$ defined in Eqs. ({\ref{Ur1}), ({\ref{Ur2}), ({\ref{Ur3}), 
({\ref{Ui1}), and ({\ref{Ui2}) to find that
\begin{eqnarray}
&& U^{\prime}_1(\v_s)=8\, U^{\prime}_2(\v_s)=-32\, U^{\prime}_3(\v_s)= -\, 
{2\epsilon_F^2\sin 4\theta \over m^2r^4} , \label{Ursingle} \\ 
&& U^{\prime\prime}_1(\v_s)=-{1\over 96}\, U^{\prime\prime}_2(\v_s) = -\, 
{\epsilon_F^2 \cos 4\theta \over 4 m^2r^4} . \label{Uisingle}
\end{eqnarray} 
Equations (\ref{dxy1}),  (\ref{Ursingle}) and (\ref{Uisingle}) together with 
the asymptotic behaviors of $L_{xy}$ in Tabel II imply that in the single vortex 
case, 
\begin{eqnarray}
L^\prime_{xy} \simeq Y^\prime (T) \left ({\xi\over r}\right )^4\sin 4\theta ,
\;\;\;\;\;\;\;\;  L^{\prime\prime}_{xy} \simeq Y^{\prime\prime} (T) 
\left ({\xi\over r}\right )^4\cos 4\theta ,  \label{lxysv}
\end{eqnarray}  
where $Y'\simeq -7.4$ in the GL regime, and $-7.1\Delta_0(T)/T$ for $T\ll 
\Delta_0$, while $Y''\simeq 9.8$ in the GL regime, and $25.9\Delta_0(T)/T$ for
$T\ll \Delta_0$. Equations (\ref{dxy1}), (\ref{lxysv}) and the asymptotic behaviors 
of $\eta_{xy}$ in Tabel II yield
\begin{eqnarray}
\Dd_{xy}^{\prime\, {\rm GL}}(\r) \simeq 3.7 \left [c_{xy}- {\Delta^2_0(T)
\over 8(\pi T)^2}+{\epsilon_Fmv^2_s(\r)\over (\pi T)^2} \right ]^{-1} 
\left ({\xi\over r}\right )^4 \sin 4\theta ,  \label{dxyrGL} \\
\bar{\D}_{xy}^{\prime\prime\, {\rm GL}}({\bf r}) \simeq -4.9 \left 
[c_{xy}-{3\over 8} {\Delta^2_0(T)\over (\pi T)^2}+{\epsilon_Fmv^2_s(\r)\over 
(\pi T)^2} \right ]^{-1} \left ({\xi\over r}\right )^4 \cos 4\theta ,
\label{dxyiGL}
\end{eqnarray}
and hence for $c_{xy}\gg 1$
\begin{eqnarray}
\Dd^{\rm GL}_{xy}(\r) \simeq - 0.61 \, i \, c_{xy}^{-1}
{\left (\xi\over r\right )^4}(e^{-i4\theta}+7e^{i4\theta}).   \label{dxyGL} 
\end{eqnarray}
This result coincides with that of Ref. \cite{Ichioka}. 

At low $T$, we obtain 
\begin{eqnarray}
\Dd^{\prime\, {\rm low} T}_{xy}(\r) &\simeq & 3.5 \left [ c_{xy} + 
{\epsilon_Fmv^2_s(\r)\over 4\Delta_0 T} \right ]^{-1} {\Delta_0\over T} 
\left ( {\xi\over r} \right )^4 \sin 4\theta,    \label{dxyrlowT}\\
\Dd^{\prime\prime\, {\rm low} T}_{xy}(\r) &\simeq &  -12.9 
\left [c_{xy}-{1\over 2}+2({\rm ln}2) {T\over \Delta_0}+{\epsilon_Fmv^2_s(\r)
\over 4\Delta_0 T} \right ]^{-1} {\Delta_0\over T} \left ( {\xi\over r} \right )^4 
\cos 4\theta,
\label{dxyilowT}
\end{eqnarray}
leading, for $c_{xy}\gg 1$, to
\begin{eqnarray}
\Dd_{xy}^{\rm low T}(\r)  
\simeq -4.69 \,i \,c_{xy}^{-1}{\left (\Delta_0\over T\right )}
{\left (\xi\over r\right )^4}(e^{-i4\theta}+1.76 \, e^{i4\theta}) .    
\label{dxylowT}
\end{eqnarray}  
In Fig. \ref{dxyvortexstructure}, we show the winding of $d_{xy}$ component and the 
magnitude of the normalized $\bar{\D}_{xy}$ at high $T$ (GL) and low 
$T$ respectively. The eight-fold symmetry is more obvious at low $T$. 

Since many current theories postulate a homogeneous $d_{x^2-y^2}+id_{xy}$ state 
without justification, it is interesting to consider the size of the spatially 
varying part of ${\cal D}_{xy}$ relative to its homogeneous component. The 
smallest relevant value of the ratio ${\cal D}_{xy}^{F_4}/{\cal D}_{xy}^{OZ}$, 
where the superscripts $F_4$ and $OZ$ indicate the relevant driving terms in 
Eqs. (\ref{dxy1}) and (\ref{outerring}), respectively, is attained at $r=R_H$ 
in the physically relevant regime $R_H<r^*$. The value is $(E_F/\Delta_0)^2 
(H/H_{c2})$ in the GL regime, and $(E_F/\Delta_0)^2 (H/H_{c2})(\Delta_0/T)$ 
at low temperatures; a simple estimate then shows that the spatially fluctuating 
component is always at least an order of magnitude larger than the homogeneous
component at experimentally relevant temperatures and fields.

\begin{figure}[h]
\begin{picture}(180,250)
\leavevmode\centering\includegraphics{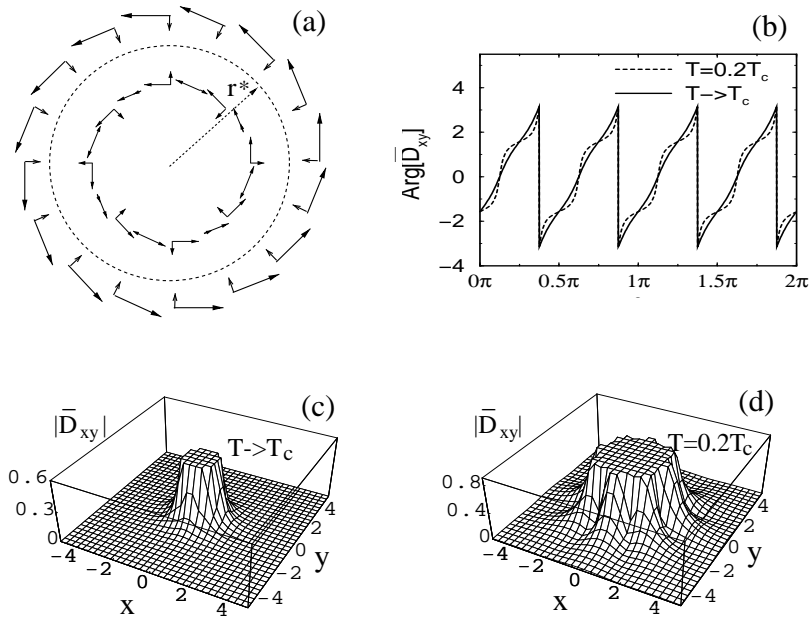}
	      \end{picture}
\caption{Single vortex structure with $d_{x^2-y^2}$ and $d_{xy}$ symmetries only. 
(a) Relative winding of $d_{xy}$ (short arrows) and $d_{x^2-y^2}$ (long arrows)
components in both the outer and inner rings. (b) Relative phase of $d_{xy}$ to 
$d_{x^2-y^2}$ component in GL and low $T$ regimes, and magnitude of the normalized 
$\bar{\D}_{xy}$ in (c) high $T$ and (d) low $T$, respectively, in the inner ring 
$\xi<r<r^*$. $c_d=3$ and $c_{xy}=7$ are chosen. Distances are given in units of $\xi$.}
\label{dxyvortexstructure}
\end{figure}

\section{Is there magnetic field-induced pseudo-phase transition?}
 
The results presented in Sections III and IV suggest that there always exist 
field-induced $s$ and $d_{xy}$ components in a parent $d_{x^2-y^2}$-wave
superconducting state where the electronic interactions in the $s$ and $d_{xy}$
channels are nonzero. These subdominant components are spatially inhomogeneous 
and, in a general case of small $V_s$ and $V_{xy}$ compared with $V_d$, are 
negligibly small as far as any bulk physical quantity is concerned. However, 
there is also a special situation in which  either $V_s$ or $V_{xy}$ is nearly 
degenerate with $V_d$, leading to a small $c_s$ or $c_{xy}$. In this case, the
denominator(s) of $\D_i, i=s,d_{xy}$ may vanish at some critical temperature 
$T^*_{ci}$, a singularity marking a second pseudo-phase transition into a 
$d_{x^2-y^2}+\D_i$ state with $\D_i$ a homogeneous bulk quantity. Investigation 
of such a possible pseudo-phase transition is particularly interesting in 
association with the experimentally observed thermal-conductivity plateau as 
mentioned in the Introduction. Since in high-$T_c$ cuprates no such phase 
transition has been reported to be found in the absence of magnetic field, we 
focus on the question whether there can be magnetic field-driven phase transition. 
This is equivalent to searching for a nonzero $T^*_{ci}$ at a finite field which 
vanishes at zero field.

We first study the $s$ component. From Eqs. (\ref{srGL}), (\ref{siGL}),
(\ref{srlowT}), and (\ref{silowT}), we see that the denominator of $\D^{\prime}_s$
can never be zero, and the singularity may occur in $\D^{\prime\prime}_s$. In the 
GL regime, the critical temperatures are 
\begin{eqnarray}
&&T^*_{cs}(\vs)={1\over \pi} \sqrt{c^{-1}_s [1/4-
2\epsilon_Fmv^2_s/\Delta^2_0(T^*_{cs})] } \Delta_0(T^*_{cs}) .    \label{Tc2GL} 
\end{eqnarray}
It follows that $T^*_{cs}(\vs)<T^*_{cs}(0)$, implying there is no instability
at nonzero field leading to a homogeneous subdominant ${\cal D}_i$. At low $T$, 
$T^*_{cs}$ becomes
\begin{eqnarray}
T^*_{cs}(\vs)={1\over 4(\,{\rm ln}\,2)} \left [1-2c_s -{\epsilon_Fmv^2_s\over 
2T^*_{cs}\Delta_0} \right ] \Delta_0 .
\label{Tc2lowTH}
\end{eqnarray} 
Again, $T^*_{cs}(\vs)<T^*_{cs}(0)$ is found, meaning that the magnetic field
does not favor such a second-order phase transition from a $d_{x^2-y^2}$ to 
$d_{x^2-y^2}+is$ state. Accounting for nonlinear effects does not affect this 
conclusion.

A similar analysis can be made in the $d_{xy}$ situation. From the asymptotic 
behaviors of $\eta^\prime_{xy}$ and $\eta^{\prime\prime}_{xy}$ listed in Table 
II, we find that there is also a singularity in $\D^{\prime}_{xy}$ in the GL 
regime, but its corresponding critical temperatures are smaller than that in 
$\D^{\prime\prime}_{xy}$. Except for this point, we reach the same conclusion
as the $s$ case: No magnetic field-induced phase transition is found.

At this stage, we may examine the prospects of a possible low-$T$ field-induced 
first-order phase transition into a time-reversal symmetry breaking state 
$d_{x^2-y^2}+id_{xy}$ as suggested by Laughlin \cite{Laughlin}. The free energy 
functional of the $d_{xy}$ pairing order parameter {\it assumed} by Laughlin 
takes the form 
\begin{eqnarray}
{F\over L^2}={1\over 6\pi}{(\D_{xy}^{\prime\prime})^3\over \nu^2}
-{1\over \pi}{eB\over c} \D_{xy}^{\prime\prime} {\rm tanh}^2\left (
{\D_{xy}^{\prime\prime}\over 2T} \right ) -{4\over \pi} {T^3\over \nu^2} 
\left \{{(\D_{xy}^{\prime\prime})^2\over 2T^2} {\rm ln} 
[1+e^{-\D^{\prime\prime}_{xy}/T}]+\int^\infty_{\D^{\prime\prime}_{xy}/T}
{\rm ln} [1+e^{-x}] x dx \right \}, \label{flaughlin}
\end{eqnarray} 
where $\nu=\sqrt{2\Delta_0/m}$. The special $T$ dependences of the second and 
third terms lead to a weak first-order phase transition at $T\simeq 0.52 
\nu \sqrt{2 eB/c}$. We may check the justification of this free energy based 
on the present theory. The first and third terms in the right-hand side of 
Eq. (\ref{flaughlin}) are understood to come from the local free energy $F^{(loc)}$ 
in Eq. (\ref{feloclau}). The second one, as the coupling of the unusual magnetization
to the field, is naturally related to the orbital Zeeman term $F^{OZ}$ in Eq. 
(\ref{OZ1}) with $M(\r)$ shown in Eq. (\ref{moment10}). But the assumed 
temperature dependence crucial for the first-order phase transition, is 
inconsistent with that of $F^{OZ}$, which we have shown to be a weak function 
of $T$ for $T\ll \Delta_0$ (see Fig. \ref{qozt}) and a linear function of 
$\D^{\prime\prime}_{xy}$ in the limit $\D^{\prime\prime}_{xy}\rightarrow 0$.
Therefore, the microscopic calculation in the present work does not confirm 
Laughlin's free energy functional which formed the basis of the first-order phase 
transition found in his work. In the present theory, a small homogenous 
$\D^{\prime\prime}_{xy}$ is found for any magnetic field and temperature in 
the superconducting state. It is swamped by a larger (but still much smaller 
than $\Delta_0$) $\D^{\prime\prime}_{xy}$ spatially fluctuating component 
over almost all of the vortex lattice for physically relevant fields.

\section{Conclusions }

In this paper, we have formulated a perturbation theory to investigate the 
magnetic-field induced subdominant order parameters of a clean $d$-wave 
superconductor in the presence of the gauge-invariant spatially varying 
supercurrent field in the mixed state. With the assumption of slowly spatially  
varying supercurrents and their induced $s$ and $d_{xy}$ components in the 
bulk sample, we are able to derive the free energy as power series in the 
Doppler energy shift, the derivatives of the supercurrents, and the subdominant
components. The free energy is valid from $T_c$ down to very low temperatures, 
enabling us to compare the resulting $s$ and $d_{xy}$ components at low $T$ 
with the existing results in the GL regime. To leading order, the real and 
imaginary parts of the $s$ component, driven by local $m(v^2_{sx}- v^2_{sy})$ 
and the derivative of the supercurrent $\partial_xv_{sx}$, respectively, were 
shown to have very different temperature dependences. In the GL regime, both 
the real and imaginary parts are in the same order. But at low $T$, the real 
part acquires an extra small prefactor $T/\Delta_0$ and the resulting $s$ 
winding happens within a very small region near antinodal directions, leaving 
a rigid $d_{x^2-y^2}+is$ state over most of the vortex. It is important to 
note, however, that this structure does {\it not} imply a gap in the 
quasiparticle spectrum, due to the small size of the $s$ component compared with 
the large Doppler shifts near the core. 

\vskip .2cm
To leading order, the $d_{xy}$ component is driven by two terms of different 
symmetries competing over different parts of the vortex lattice. The first is 
the orbital Zeeman term $F^{OZ}$ arising from the coupling of the spontaneous 
magnetization to the magnetic field. Its significance is limited by its small 
magnitude due to the particle-hole asymmetric effects. The second driving term 
$F_4$ scales as $(\xi/r)^4$ for $\xi\ll r\ll R_H$. The crossover scale $r^*$ 
divides the bulk region into inner and outer regions dominated by the two 
distinct physical effects. In the inner region, $F_4$ determines the vortex 
structure, and the $d_{xy}$ component has eight-fold symmetry and its relative 
phase winds 4 times that of the $d_{x^2-y^2}$ component. In the outer region, 
$F^{OZ}$ is more important, leading to a rigid $d_{x^2-y^2}+id_{xy}$ 
superconducting state, where the $d_{xy}$ component is spatially nearly 
homogeneous and  a weak function of temperature. We have shown that the 
crossover scale is of order $r^*=\xi (R_H/\lambda_F)^{1/2}$ in the GL regime, 
and becomes $(\Delta_0/T)^{1/4}$ times larger at low $T$. Our best estimate 
for the high-$T_c$ cuprates suggests that $r^*$ is greater than the intervortex 
separation $R_H$ for  fields above $0.01H_{c2}$, so that it appears that vortex 
lattice structure in fields of order Tesla is governed by $F_4$ and the orbital 
Zeeman effect is irrelevant. The relative phase between the $d_{x^2-y^2}$ and 
$d_{xy}$ components, as well as the magnitude of the induced $d_{xy}$, are 
therefore strongly space dependent everywhere in the sample.

\vskip .2cm
Our results have implications for several scenarios which have been proposed 
to create a state without quasiparticle excitations at low temperatures via 
the creation of a finite out-of-phase subdominant pair component. No such bulk 
state is found for generic values of the pair potentials $V_s$ and $V_d$.  It
remains possible that induced core $d_{xy}$ patches overlap with increasing 
field, as proposed by Ramakrishnan\cite{Ramakrishnan}, leading to a gapped 
state at high fields beyond the scope of our analysis. However, analyzing
the interplay between  order parameter suppression $\D_d(\r)$ around the vortex 
cores and $\v_s(\r)$, we found it  to be a very small effect on the induced 
$d_{xy}$ component; it therefore seems unlikely that the field scale where 
such a transition may occur can be significantly less than $H_{c2}$.  

\vskip .2cm
Finally, we searched for a possible second order magnetic field-induced 
phase transition for special values of the coupling constants, and reached a 
negative conclusion. No such phase transition into a bulk $d_{x^2-y^2}+s$ or 
a $d_{x^2-y^2}+d_{xy}$ state is found unless the transition has already taken 
place in the absence of the field, which is apparently not the case in the 
high-$T_c$ cuprates. Examining the Laughlin free energy driving a first-order 
phase transition, \cite{Laughlin} we found that the crucial field-dependent 
term has an assumed temperature dependence inconsistent with the BCS theory; 
thus this phase transition picture is not supported in the present work,
consistent with numerical results of Yasui and Kita. \cite{Kita}

\vskip  .2cm
The apparent phase transition observed by Krishana et al. \cite{Krishana} has 
not been clearly reproduced by other groups, and the possibility exists that 
the effect is due to inhomogeneously trapped flux. It is still interesting, 
however, to ask what kinds of intrinsic phase transitions might be possible 
in a $d$-wave superconductor.  We have shown that it is unlikely that any bulk 
phase transition  can be induced by a magnetic field, at least in the low-field 
regime where our approach is valid. Since our model neglects the vortex core 
regions, it is conceivable that vortex core transitions such as those observed 
in the $^3He$ system might still be relevant. However, since the number of 
bound states in the core region is small and possibly zero for the present 
case, it seems unlikely that such a transition would have an important effect 
on the quasiparticles responsible for heat transport at low $T$.  A final 
possibility, currently under investigation within the present framework,
is that transitions occur in the vortex lattice structure as a function of field.

\acknowledgements{The authors are grateful to N. Andrei, M. Fogelstr\"om, 
M. Franz, C. D. Gong, T. Kita, T. Kopp, S. Sachdev, S. H. Simon, and 
Y.-J. Wang for helpful communications.  Partial support was provided by the 
A. v. Humboldt Foundation, NSF Grants No. DMR-9974396 and INT-9815833, and 
``Graduiertenkolleg Anwendungen der Supraleitung'' of the Deutsche 
Forschungsgemeinschaft.}

\begin{appendix}
\section{Derivation of Free Energy}

In this Appendix we derive the free energy $\widetilde{F}$ in Eqs. 
(\ref{fetot1}-\ref{fe4}) which is valid for space region where $\v_s$ 
varies slowly.  We begin with rewriting Tr$\hat{p}^m$ in Eq. (\ref{fe1}) 
according to Eqs. (\ref{trpm}) and (\ref{vsexpansion}),
\begin{eqnarray}
{\rm Tr}\, \hat{p}^m \simeq ( {\rm Tr}\, \hat{p}^m )^{({\rm loc})} + 
( {\rm Tr}\, \hat{p}^m )^{({\rm der})} , \label{pseries}
\end{eqnarray}
where $( {\rm Tr}\, \hat{p}^m )^{({\rm loc})} $ corresponds to $\v_s(\r_j)$ 
and $ \D(\r_j)$ for all $j=1,\cdots,m-1$ taking the {\it local} values 
$\v_s(\r)$ and  $\D(\r)$, respectively, and becomes 
\begin{eqnarray}
( {\rm Tr}\, \hat{p}^m )^{({\rm loc})} = \int {d^2\k\over (2\pi)^2} 
\int d^2\r \,{\rm Tr}\, \left [\hat{p}_{\k,\k}(\r) \right ]^m ,  \label{trpm0} 
\end{eqnarray}
and $( {\rm Tr}\, \hat{p}^m )^{({\rm der})} $ contains derivatives of 
$\v_s(\r_j)$ and/or $ \D(\r_j)$. Inserting Eq. (\ref{pseries}) into Eq. 
(\ref{fe1}) leads to 
\begin{eqnarray}
F\simeq F^{({\rm loc})} +F^{({\rm der})} , 
\label{fetot}
\end{eqnarray}
where $F^{(\rm loc)}$ is just the local free energy obtained in the 
semiclassical approximation, 
\begin{eqnarray}
F^{(\rm loc)}&=& F_0-T\,{\rm Tr} \, {\rm ln} \, \widehat{M}_0 + T\,
\sum^\infty_{m=1} {1\over m} ( {\rm Tr}\, \hat{p}^m )^{({\rm loc})}
=F_0-\int d^2\r\, T\sum_n \int {d^2\k\over (2\pi)^2} 
{\rm ln} \, [-\widetilde{W}_{n\k}(\r)-\eta_\k(\r)] ,    
\label{feloc}  \\
&=& F_0-\int d^2\r \int {d^2\k\over (2\pi)^2} \sqrt{\widetilde{E}^2_\k(\r)
+ \eta_\k(\r)} - \int d^2\r \, \int {d^2\k\over (2\pi)^2} \sum_{l=\pm 1} 
{\rm ln}\left \{ 1+e^{-[l \v_s(\r) \cdot \k_F + \sqrt{\widetilde{E}^2_\k(\r)
+\eta_\k(\r)}]/T} \right \} ,  \label{feloclau}  
\end{eqnarray} 
with
\begin{eqnarray}
\widetilde{W}_{n\k}(\r)&=&-[i\omega_n+\v_s(\r) \cdot \k_F]^2
+\widetilde{E}^2_\k(\r) , \label{Wtilde}\\
\widetilde{E}_\k(\r)&=&\sqrt{\epsilon^2_\k+|\bar{\Delta}(\r)|^2 \Phi^2_{d\k}}
\nonumber \\
\eta_\k(\r)&\simeq& \sum_{i=s,d_{xy}} \left \{\left [\bar{\Delta}(\r) \D_i^*(\r)
+h.c. \right ]\Phi_{d\k} \Phi_{i\k} + |\D_i(\r)|^2\Phi_{i\k}^2 \right \} .
 \end{eqnarray}
Expanding $F^{\rm (loc)}$ in power series in $\D_i$ will give linear-$\Dd^\prime_i$
term as well as quadratic of $\D_i$ terms. The prefactors are not universal
in the whole temperature regimes, as will be discussed and shown below.  
$F^{\rm (der)}$ in Eq. (\ref{fetot}) reads
\begin{eqnarray}
F^{({\rm der})}=T\, \sum^\infty_{m=1} {1\over m} 
( {\rm Tr}\, \hat{p}^m )^{({\rm der})}=\sum_j F^{(j)},
  \label{feder}
\end{eqnarray}
where we expanded $({\rm Tr}\, \hat{p}^m)^{({\rm der})}$ as power series
in the $j$th derivatives of $\v_s$ or $\D_i$ with respect to $\r$:
$({\rm Tr}\, \hat{p}^m)^{(der)}=\sum_j ({\rm Tr}\, \hat{p}^m)^{(j)}$.
$F^{({\rm der})}$ reflects nonlocal couplings of the subdominant order 
parameters to the supercurrent fields. We examine the formal leading term, 
$F^{(1)}$ which includs $\nabla_\r v_s(\r)$ or $\nabla_\r \D_i(\r)$, in Eq. 
(\ref{feder}),  
\begin{eqnarray}
&& F^{(1)}=\sum^\infty_{m=1} {1\over m} ({\rm Tr}\, \hat{p}^m )^{(1)}
=\sum^\infty_{m=1} {1\over m} ( \rho_{m1}+\rho_{m2}+\rho_{m3}), \\
&& \;\;\; \rho_{m1}= i \sum^{m-2}_{\nu=0} \int  {d^2\k\over (2\pi)^2}\int d^2 \r
\,{\rm Tr} \left\{\,[\nabla_\r\hat{p}_{\k,\k}(\r)]\; \cdot 
\hat{p}^{m-2-\nu}_{\k,\k}(\r) \cdot [\nabla_\q\hat{p}_{\q,\k}(\r)]_{\q=\k}\;
\cdot \hat{p}^{\nu}_{\k,\k}(\r) \right \}, \\
&& \;\;\; \rho_{m2}= i \sum^{m-2}_{\nu=0} \nu \int  {d^2\k\over (2\pi)^2}
\int d^2 \r \,{\rm Tr} \left\{\,[\nabla_\r\hat{p}_{\k,\k}(\r)]\;\cdot 
\hat{p}^{m-2-\nu}_{\k,\k}(\r) \cdot [\nabla_\k\hat{p}_{\q,\k}(\r)]\;\cdot 
\hat{p}^{\nu}_{\k,\k}(\r) \right \}, \\
&& \;\;\; \rho_{m3}= i (m-1) \int  {d^2\k\over (2\pi)^2}\int d^2 \r
\,{\rm Tr} \left\{\,[(\nabla_\r\cdot\nabla_\q) \hat{p}_{\k,\q}(\r)]_{\q=\k} 
\; \cdot \hat{p}^{m-1}_{\k,\k}(\r)\right \}. 
\end{eqnarray}
$F^{(1)}$ can be expanded as power series in $\D_i$. Note in terms including 
$\nabla_\r \D_i(\r)$ the derivative can be transfered to that of $\v_s$ by 
partial integral. The resulting linear-$\D_i$ term in $F^{(1)}$ is
$F^{(1)}_\D=\sum_{i=d,s,d_{xy}} F^{(1)}_i$, where 
\begin{eqnarray}
F^{(1)}_i &\simeq &  i \int d^2 \r T\sum_n\int  {d^2\k\over (2\pi)^2} \Phi_{i\k} 
\{\nabla_\r [\v_s(\r)\cdot \k_F] \} \cdot  \sum^\infty_{m=2}
[-\v_s(\r)\cdot \k_F]^{m-2} \sum^{m-1}_{\nu=1} \nu {\rm Tr} 
\left \{ (\nabla_\k \hat{g}_\k) \cdot\hat{g}^\nu_\k \cdot\widehat{\D}_i(\r) 
\cdot\hat{g}^{m-\nu-1}_\k \right \}  \nonumber \\
&=& \int d^2 \r \,\D^{\prime\prime}_i(\r) \, T\sum_n\int {d^2\k\over (2\pi)^2}
\Phi_{i\k} 2\Delta_\k [\nabla_\k \epsilon_\k \cdot \nabla_\r] 
[\v_s(\r)\cdot \k_F]\, Z_{n\k}(\v_s) \nonumber \\
&& - \int d^2 \r \, \D^{\prime\prime}_i(\r) \,
T\sum_n\int {d^2\k\over (2\pi)^2}  \Phi_{i\k} 2\epsilon_\k [\nabla_\k\Delta_\k 
\cdot \nabla_\r] [\v_s(\r)\cdot \k_F] \, Z_{n\k}(\v_s),   \label{feder1} 
\end{eqnarray}
with $\hat{\D}_i(\r)=\left ( \begin{array}{ll} 0 & \D_i(\r) \\ 
\D^*_i(\r)  & 0 \end{array} \right )$ and 
\begin{eqnarray}
Z_{n\k}(\v_s)={\left \{[\v_s(\r)\cdot k_F]^2-W_{n\k}\right \}^2 -4 \omega^2_n 
[\v_s(\r)\cdot k_F]^2 
\over \left \{[(\v_s(\r)\cdot k_F)^2-W_{n\k}]^2+4 \omega^2_n [\v_s(\r)\cdot k_F]^2
\right \}^2 }.
\end{eqnarray}
It is easy to see that $F_{d}^{(1)}$ is negligibly small, so $F_{\bar{\Delta}}$ 
in Eq. (\ref{feH0}) is simply obtained from $F^{(loc)}$ in Eq. (\ref{feloc}) by 
setting $\D_s, \D_{xy}=0$. We argue that $F^{(loc)}$ and $F^{(1)}_s$ in Eqs. 
(\ref{feloc}) and (\ref{feder1}) are sufficient for studying the $s$ component 
up to leading orders. Expanding $F^{(loc)}$ as power series in $\D_s$ and $v_sk_F$,
we will see that the driving term for $\D^{\prime}_s$ is in leading order of
$m(v^2_{sx}-v^2_{sy})$, which, in the spatial regime of interest $\xi<r<R_H$, 
scales as $1/r^2$. $F^{(1)}_s$ gives a driving term for $\D^{\prime\prime}_s$ 
scaling as $\partial_x v_{sx}\sim 1/r^2$.  Clearly, terms contained in 
$\sum_{l>1}F^{(l)}$ are of higher orders of $v_s$ and derivatives of $v_s$. 
We first show results for $s$ component. The $d_{xy}$ situation is more 
complicated and will be discussed later.

\subsection{Scaling expressions for $L_s(T,\v_s)$ and $\eta_s(T,\v_s)$ 
at $T>\v_s\cdot\k_F$}
 
In this temperature regime, we can expand $\widetilde{W}_{n\k}(\r)$ in Eqs. 
(\ref{feloc}) and (\ref{feder1}) as power series in $v_sk_F$, $\D^{\prime}_i(\r)$, 
and $\D^{\prime\prime}_i(\r)$ to find explicit expressions for $L_s(T,\v_s)$ and 
$\eta_s(T,\v_s)$ in Eq. (\ref{fetots}), 
\begin{eqnarray}
&& L_s'(T,\v_s) = N^{-1}_0 T\sum_n \int {d^2\k \over (2\pi)^2} 
{2\cos^2 2\varphi (4\omega^2_n-W_{n\k}) \over W^3_{n\k} }\epsilon_F m 
[v^2_{sx}-v^2_{sy}] =2{\epsilon_F m [v^2_{sx}-v^2_{sy}] \over T^2} h_{1s}
\left (d \right ),    \label{Qsr1} \\
&&  L_s^{\prime\prime}(T,\v_s) = T\sum_n \int {d^2\k \over (2\pi)^2} 
{4\cos^2 2\varphi \over W^2_{n\k} } \epsilon_F \partial_x v_{sx}
= 4{\epsilon_F \partial_x v_{sx}\over T^2} h_{2s}\left (d \right ), 
\label{Qsi1} \\
&& \eta^\prime_s(T,\v_s) =\eta^{\prime\prime}_s(T,\r)+2d^2 h_{2s}(d) , 
\label{Qsr2} \\
&&\eta^{\prime\prime}_s(T,\v_s) = -h_{3s}(d)+{\epsilon_F mv^2_{s}(\r) 
\over T^2} h_{4s}(d),  \label{Qsi2}  
\end{eqnarray}
where Eq. (\ref{dgapH0}) was used, $d=\Delta_0(T)/T$, and
\begin{eqnarray}
&&h_{1i}(d)={8\over \pi}\int^\infty_0 dx \int^{\pi/4}_0 d\varphi 
\cos^2 2\varphi\, \Phi_{i\k}^2 {1\over 2\eta}{d^2f(\eta)\over d\eta^2} , 
\;\;\;\;\;\;\;\;\;\;\;\;\;\;
h_{2i}(d)={8\over \pi}\int^\infty_0 dx \int^{\pi/4}_0 d\varphi 
\cos^2 2\varphi \, \Phi_{i\k}^2 {t(\eta)\over 4\eta^3}, \label{h12i} \\
&&h_{3i}(d)={8\over \pi}\int^\infty_0 dx \int^{\pi/4}_0 d\varphi 
(1-2\cos^2 2\varphi) \,\Phi_{i\k}^2 {{\rm tanh}(\eta/2)\over 2 \eta}, 
\;\;\;\;\;\;\;\;\;\;\;\;\;\;
h_{4i}(d)={8\over \pi}\int^\infty_0 dx \int^{\pi/4}_0 d\varphi \, 
\Phi_{i\k}^2 {1\over 2\eta}{d^2f(\eta)\over d\eta^2} , \label{h34i} 
\end{eqnarray} 
with $\eta=\sqrt{x^2+d^2\cos^2 2\varphi}$, $f(\eta)=1/(e^\eta+1)$,  
and $t(\eta)=1-2f(\eta)+2\eta df(\eta)/d\eta$.

\subsection{ $L_s(T,\v_s)$ and $\eta_s(T,\v_s)$ at $T=0$}

At extremely low $T$, the perturbation expansions of the free energy as 
power series in $v_s(\r)k_F$ and $\D^{\prime}_i(\r)$ by expanding 
$\widetilde{W}_{n\k}(\r)$ {\it before} integrating over $\k$ breaks down 
due to the existence of the zero modes in $E_\k$. The correct approach is 
to expand the free energy {\it after} doing the integrals over $\k$. For 
simplicity, we only show the $T=0$ results. It is easy to find that the 
driving term for $\D^{\prime}_s$ is negligibly small compared with that 
for $\D^{\prime\prime}_s$. We may simply set $\D^{\prime}_s=0$. The local 
free energy in Eq. (\ref{feloc}) becomes
\begin{eqnarray}
F^{(loc)}_{T=0}=F_0- \int d^2\r \int {d^2\k\over (2 \pi)^2 } 
\sqrt{\widetilde{E}^2_\k(\r)+\eta_\k(\r) }+2 \int d^2\r 
\int {d^2\k\over (2 \pi)^2 } Y_\k(\r)  \Theta (-Y_\k(\r) )
\end{eqnarray}
where $Y_\k(\r)=\sqrt{\widetilde{E}^2_\k(\r)+[\D^{\prime\prime}_s(\r)]^2}
-\v_s(\r)\cdot\k_F$ and $\Theta(x)$ is the Heaviside step function.  
We obtain $F^{(loc)}_{T=0}=\Delta F+\int d^2\r [c_s +\eta_s''(T=0,\v_s)] 
[\Dd^{\prime\prime}_s(\r)]^2$, 
where $\Delta F$ is the local free energy at $\D_s=0$, and
\begin{eqnarray}
&& \eta_s''(T=0,\v_s) \simeq  -{1\over 2} + S_\theta {v_s(\r)k_F\over 2\Delta_0}, 
\label{Qsi2T0} 
\end{eqnarray}
with $S_\theta=\sum_{l=\pm 1} \left |\cos (\theta+l {\pi\over 4}) \right |$. 

From $F^{(1)}_s$ in Eq. (\ref{feder1}) we get the driving term for 
$\D^{\prime\prime}_s$, from which we find 
\begin{eqnarray}
&& L_s''(T=0,\v_s) \simeq {2 \epsilon_F m [v^2_{sx}-v^2_{sy}]\over\Delta^2_0} 
\left [1-S_\theta {v_s(\r)k_F\over \Delta_0} \right ].  \label{Qsi1T0}
\end{eqnarray}

\subsection{Free energy with respect to $d_{xy}$ component}
 
We first analyze the driving terms at $T>\v_s\cdot\k_F$. The leading-order 
linear-$\D_{xy}$ term from $F^{(loc)}$ in Eq. (\ref{feloc}) turns out to be 
\begin{eqnarray}
F^{(loc)}_4\simeq \int d^2\r \Delta^2_0(T) N_0 Q^{\prime}_3(T) U^{\prime}_3
(\v_s) \bar{\D}^\prime_{xy}, \label{fe4loc}
\end{eqnarray}
where 
\begin{eqnarray}
Q^{\prime}_3(T)&=& N^{-1}_0 T\sum_n \int {d^2\k\over (2\pi)^2}  \sin^2 4\varphi
\left \{ -{5\over 2 W^3_{n\k}}+{10 E^2_\k \over W^4_{n\k}}-{8 E^4_\k \over 
W^5_{n\k}} \right \} ,  \label{Q43r} \\
U^{\prime}_3(\v_s)&=&4\epsilon_F^2m^2\left [v_{sx}^2(\r)-v_{sy}^2(\r)\right ]
v_{sx}(\r)v_{sy}(\r) . \label{Ur3}
\end{eqnarray}
For $\xi<r<R_H$, $F^{(loc)}_4$ is of order  $1/r^4$. As for $F^{(1)}_{xy}$ 
in Eq. (\ref{feder1}), the two terms in the right-hand side are qualitatively 
different. We leave the discussion on the second term for a bit later. The first 
term is nonzero in the leading order of $(mv^2_s)\partial_xv_{sx}$ which scales 
as $1/r^4$ too. For $T>v_sk_F$, this term is
\begin{eqnarray}
F^{(11)}_{xy}\simeq \int d^2\r\Delta^2_0(T) N_0 Q^{\prime\prime}_1(T) 
U^{\prime\prime}_1(\v_s) \, \bar{\D}^{\prime\prime}_{xy},
\end{eqnarray}
where 
\begin{eqnarray}
Q^{\prime\prime}_1(T)&=& N^{-1}_0 T\sum_n \int {d^2\k\over (2\pi)^2} 
\sin^2 4\varphi \left \{ -{5\over W^3_{n\k}}+{6 E^2_\k \over W^4_{n\k}} 
\right \}, \label{Q41i} \\
U^{\prime\prime}_1(\v_s)&=&\epsilon_F^2\left \{ m \left [v_{sx}^2(\r)
-v_{sy}^2(\r) \right ] \left [ \partial_x v_{sy}(\r)+ \partial_y v_{sx}(\r) 
\right ] + 4 m v_{sx}(\r)v_{sy}(\r) \partial_x v_{sx}(\r) \right \} . \label{Ui1} 
\end{eqnarray}
A simple analysis shows that there are terms coming from $F^{(2)}$ and 
$F^{(3)}$ in Eq. (\ref{feder}) which are of the same order. After some algebra, 
we find that the terms of order $1/r^4$ in $F^{(2)}$ are
\begin{eqnarray}
F^{(2)}_{xy}&\simeq &\int d^2\r  \Delta^2_0(T) N_0\; \left \{ Q^{\prime}_1(T) 
U^{\prime}_1(\v_s) + Q^{\prime}_2(T)U^{\prime}_2(\v_s) \right \} 
\bar{\D}^\prime_{xy}, \label{fe2dxy} 
\end{eqnarray}
where 
\begin{eqnarray}
Q^{\prime}_1(T)&=& N^{-1}_0 T\sum_n \int {d^2\k\over (2\pi)^2}  \sin^2 4\varphi
\left \{ {3\over W^3_{n\k}}-{4 (E^2_\k+3\epsilon_\k^2)\over W^4_{n\k}}
+{16 \epsilon_\k^2 E^2_\k  \over W^5_{n\k}}   \right \},   \label{Q41r}  \\
Q^{\prime}_2(T) &=& N^{-1}_0 T\sum_n \int {d^2\k\over (2\pi)^2} \sin^2 4\varphi 
\left \{ {2\over W^3_{n\k}}+{24\epsilon_\k^2 \over W^4_{n\k}}
- {32\epsilon_\k^2 E^2_\k \over W^5_{n\k}}  \right \} , \label{Q42r}  \\
U^{\prime}_1(\v_s) &=& \epsilon_F^2\left \{ v_{sx}(\r) \left [\partial^2_x v_{sy}(\r) +
3 \partial_x \partial_y v_{sx}(\r) \right ]  
+ v_{sy}(\r) \left [ -\partial^2_y v_{sx}(\r)
+3 \partial^2_x  v_{sx}(\r) \right ] \right \} , \label{Ur1}\\
U^{\prime}_2(\v_s) &=& \epsilon_F^2[\partial_x v_{sx}(\r)] [\partial_x v_{sy}(\r) + 
\partial_y v_{sx}(\r)], \label{Ur2}
\end{eqnarray}
and in $F^{(3)}$ are
\begin{eqnarray}
F^{(3)}_{xy}&\simeq &\int d^2\r \Delta^2_0(T) N_0 Q^{\prime\prime}_2(T) 
U^{\prime\prime}_2(\v_s) \bar{\D}^{\prime\prime}_{xy} , \label{fe3dxy} 
\end{eqnarray}
where
\begin{eqnarray}
Q^{\prime\prime}_2(T)&=& N^{-1}_0 T\sum_n \int {d^2\k\over (2\pi)^2} \sin^2 4\varphi 
\left \{ {1\over 2W^3_{n\k}}-{2\epsilon_\k^2 \over W^4_{n\k}} \right \} , 
\label{Q42i} \\
U^{\prime\prime}_2(\v_s) &=&\epsilon_F^2 \left [{1\over m} \partial^3_x v_{sy}(\r)
-{1\over m} \partial^3_y v_{sx}(\r) + {6\over m}\partial^2_x \partial_y v_{sx}(\r) 
\right ]. \label{Ui2}
\end{eqnarray}
The sum of $F^{(loc)}_4, F^{(11)}_{xy}, F^{(2)}_{xy}$, and $F^{(3)}_{xy}$ leads 
to $F_4(\r)$ term in Eq. (\ref{fe4}), with 
\begin{eqnarray}
L^{\prime}_{xy}(T,\v_s) &=& Q^{\prime}_1(T) U^{\prime}_1(\v_s) + 
Q^{\prime}_2(T)U^{\prime}_2(\v_s) + Q^{\prime}_3(T)U^{\prime}_3(\v_s) ,  
\label{lrxy} \\ 
L^{\prime\prime}_{xy}(T,\v_s) &=& Q^{\prime\prime}_1(T) 
U^{\prime\prime}_1(\v_s) +Q^{\prime\prime}_2(T) 
U^{\prime\prime}_2(\v_s) . \label{lixy}
\end{eqnarray}

The second term on the right-hand side of Eq. (\ref{feder1}) is nonvanishing 
by noting that $[\nabla_\k\Delta_\k \cdot \nabla_\r] [\v_s(\r)\cdot k_F]=\Delta_0 
\sin 2\varphi [\nabla_\r \times \v_s(\r)]_z$ which picks up the magnetic vector 
potential $\A(\r)$ in $\v_s(\r)$. This term is nothing but $F^{OZ}$ in Eq. 
(\ref{feOZ}), with 
\begin{eqnarray}
Q^{OZ}(T)=- N^{-1}_0 T\sum_n \int {d^2\k\over (2\pi)^2}  
2 \epsilon_\k \sin^2 2\varphi Z_{n\k}(\v_s)
\simeq - N^{-1}_0 T\sum_n \int {d^2\k\over (2\pi)^2}  
{2 \epsilon_\k \sin^2 2\varphi \over W^2_{n\k} } .  \label{qoz}
\end{eqnarray}
$F^{OZ}$ is very different from $F_4$. Discussions about comparative importance 
of these two terms can be found in Sec. III.C.  The integrand of $Q^{OZ}$
includes odd power in $\epsilon_\k$, implying nonzero contribution only in a 
particle-hole asymmetric system. In order to proceed further, we adopt the 
following density of states near the Fermi surface in the normal state,
\begin{eqnarray}
N(\epsilon)\simeq N_0+\epsilon \left. {dN(\epsilon)\over d\epsilon} 
\right |_{\epsilon=0} \simeq N_0 (1+\alpha {\epsilon \over \epsilon_F}), 
\label{DOS}
\end{eqnarray}
to take into account both the particle-hole symmetric and asymmetric contributions.
Here $\alpha$ is of order unity, and $\epsilon$ is typically order of $T$ or 
$\Delta_0$, implying that the second term on the right-hand side of Eq. (\ref{DOS}) 
may be negligible when the contribution from the particle-hole symmetric mode does 
not vanish. 

The quadratic free energy terms in Eq. (\ref{fetotdxy}) are extracted from 
$F^{(loc)}$, where 
\begin{eqnarray}
&& \eta^{\prime}_{xy}(T,\v_s) =\eta^{\prime\prime}_{xy}(T,\v_s) 
+ 2d^2 h_{2d{xy}}(d), \label{Qdxyr2} \\
&& \eta^{\prime\prime}_{xy}(T,\v_s) = -h_{3xy}(d)+{\epsilon_F mv^2_{sx}(\r) 
\over T^2} h_{4xy}(d) ,  \label{Qdxyi2}  
\end{eqnarray}
with $h_{id{xy}}$ defined in Eqs. (\ref{h12i}) and (\ref{h34i}).

\section{Spontaneous magnetization in the \protect{$d_{x^2-y^2}+id_{xy}$} state}

To proceed with a general derivation of the spontaneous 
magnetization, we rewrite the free energy in Eq. (\ref{fe1}) as
\begin{eqnarray}
&&F = F_0 - T\sum_n {\rm Tr} \, {\rm ln}  
\left (-\widehat{G}^{-1} \right ) - T\sum_n {\rm Tr} \, {\rm ln}  
\left [ 1- \widehat{G} 
\left (     \begin{array}{cc}
\widehat{V}_1  & 0  \\
0   &  \widehat{V}_2 
  \end{array}
\right ) \right ], \label{F1}
\end{eqnarray}
where $\widehat{G}= -(\widehat{M}_0+\widehat{D})^{-1}$ with $\widehat{D}$ the 
field-induced off-diagonal pairing order parameter matrix. The orbital Zeeman 
term coming from the linear-$\widehat{V}$ term is
\begin{eqnarray}
F^{OZ} &\simeq & - T\sum_n \int {d^2\k d^2\k_1 \over (2\pi)^4} 
\int d^2\r d^2 \r_1 e^{i(\k_1-\k)\cdot (\r_1-\r)} \left [G_{11 \k,\k_1}(\r) 
+ G_{22 \k,\k_1}(\r) \right ] \left [ (\r_1-\r) \cdot \nabla_\r \right ] 
[\v_s(\r)\cdot \k_F ] \nonumber\\
&\simeq & T\sum_n \int {d^2\k \over (2\pi)^2} \int d^2\r 
{\rm lim}_{\k_1\rightarrow\k} \left \{ i\nabla_\k \left [ G_{11 \k,\k_1}(\r) + 
G_{22 \k,\k_1}(\r) \right ] \cdot\nabla_\r \right \} [\v_s(\r)\cdot \k_F ]  
\nonumber \\
& \simeq & -\int d^2\r {e\over 2mc} T \sum_n \int {d^2\k \over (2\pi)^2}  
{\rm lim}_{\k_1\rightarrow\k} \left \{ (i\k\times \nabla_\k) 
\left [ G_{11 \k,\k_1}(\r) + G_{22 \k,\k_1}(\r) \right ] \right \}
\cdot {\bf B}(\r).  \label{feOZ0} 
\end{eqnarray} 
Comparing Eq. (\ref{feOZ0}) with Eq. (\ref{OZ1}) we see that the magnetization 
is just the expectation value of the magnetic moment operator $(e/2mc) 
(i\k\times \nabla_\k)$, 
\begin{eqnarray}
{\bf M}(\r)={e\over 2mc}  \int {d^2\k \over (2\pi)^2} {\rm Tr}\langle 
\hat{d}^\dagger_\k (i\k\times \nabla_\k) \hat{d}_\k \rangle 
={e\over 2mc} T\sum_{n}\int {d^2\k \over (2\pi)^2} {\rm lim}_{\k_1\rightarrow \k}
(i\k\times \nabla_\k) {\rm Tr} \widehat{G}_{\k,\k_1}(\r) ,
\label{magmoment}
\end{eqnarray}
with $\hat{d}^\dagger_\k=(c_{\k\uparrow},c^\dagger_{-\k\downarrow})$.
It is easy to check that ${\bf M}=0$ for a pure $d_{x^2-y^2}$- ($d_{xy}$-)
wave superconductor, and also for a $d_{x^2-y^2}+is$-wave one in the order 
of linear $s$. For a $d_{x^2-y^2}+id_{xy}$-wave superconductor 
with $d_{xy}$ component perturbatively small compared to the $d_{x^2-y^2}$ 
component, we use the Dyson equation,
\begin{eqnarray}
\widehat{G}_{\k,\k_1}\simeq \hat{g}_\k \delta(\k-\k_1) + \hat{g}_\k 
\left (    \begin{array}{cc} 0  &  \D_{xy}(\r) \Phi_{xy{\k+\k_1\over 2}} \\
\D^*_{xy}(\r) \Phi_{xy{\k+\k_1\over 2}}  &  0 
\end{array}  \right )
\hat{g}_{\k_1},  \label{GFk}
\end{eqnarray}
where $\hat{g}_\k$ is defined in Eq. (\ref{gf}). Inserting Eq. (\ref{GFk}) 
into Eq. (\ref{magmoment}) yields
\begin{eqnarray}
{\bf M}(\r) &=& {e\over 2mc}  iT \sum_{n}\int {d^2\k \over (2\pi)^2}\Phi_{xy\k} 
\left [ \D_{xy}(\r) g_{4\k} + \D^*_{xy}(\r) g_{1\k} \right ] 
(\k\times \nabla_\k) g_{2\k} \label{moment} \\
&=& -{e\over 2mc} T \sum_{n}\int {d^2\k \over (2\pi)^2} \D^{\prime\prime}_{xy}(\r)
{4\Delta_0 \epsilon_\k \sin^2 2\varphi \over W^2_{n\k} }, 
\end{eqnarray}
which, up to the leading order of $D_{xy}$ and $v_s$, is consistent with the 
second term on the right-hand side of Eq. (\ref{feder1}).

A particle-hole asymmetric system is required to obtain a nonvanishing
${\bf M}(\r)$. To understand the physics, we note ${\bf M}(\r)$ in Eq. 
(\ref{moment}) is in fact 
\begin{eqnarray}
{\bf M}(\r)&=&{e\over 2mc} \sum_{n,\k} 2 \D^{\prime\prime}_{xy} 
 \left \{(u^2_\k-v^2_\k) {1\over i\omega_n+E_\k} - (u^2_\k-v^2_\k) 
 {1\over i\omega_n-E_\k} \right \} (\k \times \nabla_\k) g_{2\k} .   \label{mag}
\end{eqnarray}
with $u^2_\k=(1+\epsilon_k/E_\k)/2$ and $v^2_\k=(1-\epsilon_k/E_\k)/2$ measuring 
the particle and hole populations, respectively. Eq. (\ref{mag}) can be 
interpreted as a magnetic moment contributed from number currents of both 
particles and holes flowing in the opposite directions, which cancel in a 
particle-hole symmetric system.

The spontaneous magnetization $M(\r)$ as a full expression of 
$\bar{\D}^{\prime\prime}_{xy}$ can also be obtained from Eq. (\ref{magmoment}).
After some straightforward algebra we get 
\begin{eqnarray}
{\bf M}(\r)\simeq -{eT\over 2mc}  \sum_{n}\int {d^2\k \over (2\pi)^2} 
{4\Delta_0(T) \D^{\prime\prime}_{xy}(\r)\epsilon_\k \Phi^2_{xy\k}\over W_{n\k} 
\{W_{n\k}+ [\D^{\prime\prime}_{xy}(\r)]^2\Phi_{xy\k}^2 \}  } . \label{moment10}
\end{eqnarray}
${\bf M}(\r)$ in Eq. (\ref{moment10}) is important for the direct comparison with 
Laughlin's free energy as discussed in Sec. V.

\end{appendix}


\begin{references}
\bibitem{gaugetran} We use a symmetric gauge transformation criticized by Franz 
and Tesanovic \cite{FranzTes} which we believe to be justified in the current
application. This follows because the free energy involves only the energy 
eigenvalues, which do not depend on the choice of gauge.
\bibitem{footnote4} We note that the second term $\sim [\bar{\cal D}_i(\r)^2 -
\bar{\cal D}_i^*(\r)^2]$ is allowed by symmetry, but was neglected
in the GL analysis of the collective mode in the $d_{x^2-y^2}+id_{xy}$ state
of Ref. \cite{BalatskyKumar}, and may in principle substantially change the 
collective mode frequency calculated therein.\cite{Sachdev}  Within the weak 
coupling analysis presented here, the prefactor is of order $c^{-1}_{xy}$, 
which is small for generic values of the coupling constants $V_d,V_{xy}$.
Therefore, the neglect of this term in Ref. \cite{BalatskyKumar} is generally 
justified within weak-coupling theory, except in cases of near degeneracy of
the $d_{x^2-y^2}$, $d_{xy}$ coupling constants.
\bibitem{note1} In a vortex crystal, however, the quasiparticles take the form 
of Bloch waves in the presence of the periodic supercurrents; thus the 
perturbation theory presented in the present paper breaks down. Even in this 
case, however, the semiclassical approach we have adopted may yield sensible 
results.\cite{Melnikov}
\end{references}
\end{document}